\documentclass[preprint2,longabstract]{aastex}

\received{ 8 July 2005} \revised{28 April 2006} \accepted{1 May
2006}

\slugcomment{To appear in Astrophys. J.} \shorttitle{Different ISM
Environments of the Heliosphere} \shortauthors{M\"uller et al.}

\newcommand {\gtsim} {\ {\raise-.5ex\hbox{$\buildrel>\over\sim$}}\ }
\newcommand {\ltsim} {\ {\raise-.5ex\hbox{$\buildrel<\over\sim$}}\ }
\newcommand{\Vlsr}{V$_{\rm LSR }$}
\newcommand{\deeg}{$^{\rm o }$}
\newcommand{\cc}{cm$^{-3}$}
\newcommand{\glong}{$l$}
\newcommand{\glat}{$b$}
\newcommand{\ce}{charge exchange}
\newcommand{\kms}{km s$^{-1}$}
\newcommand{\HI}{H$^{\rm o}$}
\newcommand{\HII}{H$^{\rm +}$}

\newcommand{\nHI}{$n$(\HI)}
\newcommand{\nH}{$n$(\HI)}
\newcommand{\np}{$n$(\HII)}
\newcommand{\nel}{$n$(e$^{\rm - }$)}
\newcommand{\NHI}{$N$(\HI)}
\newcommand{\NCaII}{$N$(Ca$^{\rm + }$)}
\newcommand{\NH}{$N$(H)}

\newcommand{\NFeII}{$N$(Fe$^{\rm + }$)}
\newcommand{\CaII}{Ca$^{\rm + }$}

\newcommand{\DI}{D$^{\rm o }$}

\newcommand{\NDI}{$N$(D$^{\rm o }$)}

\newcommand{\lya}{Ly-$\alpha$}
\newcommand{\cmtwo}{cm$^{-2}$}
\newcommand{\ntot}{$n_{\rm tot}$}
\newcommand{\chiH}{$\chi$(H)}

\newcommand{\nummodels}{27}
\newcommand{\prlocbub}{1}
\newcommand{\hdsub}{2}
\newcommand{\prtwo}{3}
\newcommand{\preight}{4}
\newcommand{\prtwelve}{5}
\newcommand{\prsixt}{6}
\newcommand{\preleven}{7}
\newcommand{\prfift}{8}
\newcommand{\acen}{9}
\newcommand{\azz}{10}
\newcommand{\prone}{11}
\newcommand{\prseven}{12}
\newcommand{\azzw}{13}
\newcommand{\hivel}{14}
\newcommand{\hidensa}{15}
\newcommand{\hidensc}{16}
\newcommand{\hidensb}{17}
\newcommand{\xiboo}{18}
\newcommand{\oph}{19}
\newcommand{\evlac}{20}
\newcommand{\vir}{21}
\newcommand{\eind}{22}
\newcommand{\eindmod}{23}
\newcommand{\eindmodm}{24}
\newcommand{\cyg}{25}
\newcommand{\colda}{26}
\newcommand{\coldb}{27}

\begin{document}

\title{Heliospheric Response to Different Possible Interstellar Environments}

\author{Hans-Reinhard M\"{u}ller\altaffilmark{1}}
\affil{Department of Physics and Astronomy, Dartmouth College,
Hanover, NH 03755.} \altaffiltext{1}{also at: Institute of
Geophysics and Planetary Physics, University of California,
Riverside, CA 92521.} \email{hans.mueller@dartmouth.edu}

\author{Priscilla C. Frisch}
\affil{Department of Astronomy and Astrophysics, University of
Chicago, Chicago, IL 60637.} \email{frisch@oddjob.uchicago.edu}

\author{Vladimir Florinski and Gary P. Zank}
\affil{Institute of Geophysics and Planetary Physics, University
of California, Riverside, CA 92521.}
\email{vladimir.florinski@ucr.edu, gary.zank@ucr.edu}

\begin{abstract}
At present, the heliosphere is embedded in a warm low density
interstellar cloud that belongs to a cloud system flowing through
the local standard of rest with a velocity near $\sim$18 \kms. The
velocity structure of the nearest interstellar material (ISM),
combined with theoretical models of the local interstellar cloud
(LIC), suggest that the Sun passes through cloudlets on timescales
of $\le 10^3$--10$^4$ yr, so the heliosphere has been, and will
be, exposed to different interstellar environments over time. By
means of a multi-fluid model that treats plasma and neutral
hydrogen self-consistently, the interaction of the solar wind with
a variety of partially ionized ISM is investigated, with the focus
on low density cloudlets such as are currently near the Sun. Under
the assumption that the basic solar wind parameters remain/were as
they are today, a range of ISM parameters (from cold neutral to
hot ionized, with various densities and velocities) is considered.
In response to different interstellar boundary conditions, the
heliospheric size and structure change, as does the abundance of
interstellar and secondary neutrals in the inner heliosphere, and
the cosmic ray level in the vicinity of Earth. Some empirical
relations between interstellar parameters and heliospheric
boundary locations, as well as neutral densities, are extracted
from the models.
\end{abstract}


\section{INTRODUCTION}

The heliosphere is a low density cavity that is carved out from
the local interstellar medium (LISM) by the solar wind. The size
and particle content of the heliosphere are determined by the
solar wind -- LISM interaction, and they vary in response to the
Galactic environment of the Sun as the Sun and interstellar clouds
move through space. The path of the Sun has taken us through the
Local Bubble void \citep[galactic longitudes
180\deeg$\ltsim$\glong$\ltsim$270\deeg,][]{FrischYork:1986}, and
we have recently ($\ltsim$ 10$^3 - 10^5$ yr ago, depending on
cloud shapes and densities) entered a clumpy flow of low density
interstellar material \citep[][]{Frisch94}. This clumpy flow, the
``cluster of local interstellar cloudlets'' (CLIC), is flowing
away from the Sco-Cen association and extends 10--30 pc into the
Galactic center hemisphere and $\ltsim$3 pc for many directions in
the anticenter hemisphere.  Inhomogeneities in the CLIC create
temporal variations in the dynamic interstellar pressure at the
solar location, which may produce significant variations in
heliosphere properties over geologically short timescales
\citep{Frisch:1993a,Frisch:1997,ZankFrisch99,Florinskietal03a,Frischetal:2002,Frisch:2004igpp}.

The heliosphere itself is a dynamically changing object which is
highly sensitive to interstellar pressure
\citep[e.g.][]{Holzer:1989,Zank99}.  The interaction of the ISM
with the fully ionized solar wind gives rise to the heliospheric
morphology which includes the heliopause (HP), a tangential
discontinuity separating solar wind and LISM, and the termination
shock (TS) where the solar wind becomes subsonic and is diverted
downstream to form a heliotail.  Depending on the pressure of the
surrounding interstellar material, an interstellar bow shock (BS)
may form upwind of the heliopause. These general boundaries are
created by the plasma interaction, yet the presence of neutral H
and its coupling to the plasma protons via charge exchange greatly
influences the details of the heliospheric morphology and the
location of its boundaries (see \citet{Zank99} for a review). The
sensitivity of the heliosphere to variations in the physical
characteristics of the interstellar cloud surrounding the solar
system is poorly understood, and in this paper we focus on
heliosphere variations due to encounters with a range of low
density clouds such as expected in the immediate past and future
solar history.

Different interstellar environments may produce noticeable changes
in the interplanetary environment of the inner heliosphere, as
indicated by the amount of neutral H, anomalous, and galactic
cosmic rays (GCR) at 1 AU.  There is some evidence that lunar
soils contain an archive of isotopic abundances that are different
from the particle environment of the present era \citep{rfw00},
and antarctic ice cores show signatures that may be interpreted as
cosmic ray background variations at Earth
\citep{Be10,SonettJokipii:1987,FlorinskiZank05}.  These
possibilities have motivated our study of the behavior of the
global heliosphere under variable boundary conditions resulting
from passage through interstellar clouds.

Given the inhomogeneity of the local solar neighborhood and the
galactic environment in general, we test the heliosphere response
to a range of local interstellar boundary parameters using about
two dozen specific parameter sets. Our choices are justified in \S
\ref{ss:nearbyism}. Four highlights of the corresponding
heliospheric models are detailed in \S \ref{ss:exmodels}. The
results of all the heliospheric models calculated for this study
suggest relationships of the heliospheric boundary locations and
the neutral particle densities with the interstellar parameters,
discussed in \S \ref{ss:results}. The synopsis of all the
individual model results through these relations is a quantitative
expression of the sensitivity of the heliosphere to changing
interstellar boundary conditions. Furthermore, the relations allow
for a prediction of boundary locations and particle content for
heliospheres with yet different boundary parameters, without
actually engaging in a complex, non-linear global heliosphere
simulation. This predictive power can also be used in the emerging
field of astrospheres, which are the analogues of heliospheres
around solar-like cool stars.

We discuss the response of the global heliosphere to variable
interstellar properties, and speculate on aspects of the
implications of these variations for the 1 AU location of the
Earth in \S \ref{ss:discuss}.

\section{PROPERTIES OF ISM IN THE SOLAR NEIGHBORHOOD  \label{ss:nearbyism}}

The Sun is embedded in a flow of warm low density gas with an
upwind direction in the local standard of rest directed towards
the Scorpius-Centaurus Association \citep{Frisch:2004igpp}.  The
CLIC is defined by high resolution absorption lines towards nearby
stars, and lower resolution observations of white dwarf stars in
the far ultraviolet (UV) and extreme UV.  With the possible
exception of $\alpha$ Oph, interstellar column densities towards
stars within 35 pc of the Sun do not exceed $\sim$10$^{19}$
\cmtwo\ \citep[e.g.][]{Frisch:2004igpp,RLII,Woodetal:2005}. To
date, over 150 absorption components have been identified by
velocity in optical and UV observations of at least 90 stars
sampling the nearby ISM \citep[see references
in][]{FGW:2002,RLI,RLII,RLIII}. The ISM towards nearby white dwarf
stars is partially ionized
\citep{Vallerga:1998,Holbergetal:1999,Frisch:2004igpp}, and local
variations in \NFeII/\NDI\ show that the CLIC is inhomogeneous
\citep{Frisch:2004igpp}.  From these data, a general picture has
emerged that the ISM within $\sim$35 pc of the Sun is dominated by
low density warm gas.  The general properties of this nearby ISM
are consistent with partially ionized, low column density gas (log
\NHI\ $< 18$ dex, $N$ in \cmtwo) described by radiative transfer
models \citep{SlavinFrisch:2002}.


\subsection{Short-term Variations in the Solar Environment
\label{ss:shortterm}}

The properties of the CLIC are diagnosed by Doppler-broadened
absorption features representing clouds (or ``cloudlets'')
observed in the optical and UV data towards $\sim$100 nearby
stars.  The best-fitting flow velocity in the Local Standard of
Rest (LSR) is $-19.4\pm$4.6 \kms, with an upwind direction \glong\
= 331.4\deeg, \glat\ = --4.9\deeg. For comparison, the LSR local
interstellar cloud (LIC) velocity is --20.6 \kms, and the upwind
direction is (\glong, \glat) = (317.8\deeg, --0.5\deeg). These LSR
values assume the standard solar apex motion of 19.7 \kms\ towards
\glong\ = 57\deeg, \glat\ = +22\deeg. The corresponding
heliocentric flow vector is --28.1$\pm$4.6 \kms\ from the upwind
direction (\glong, \glat) = (12.4\deeg, 11.6\deeg).\footnote{An
alternate solar apex motion, based on Hipparcos data \citep[13.4
\kms\ towards \glong\ = 27.7\deeg, \glat\ =
32.4\deeg,][]{DehnenBinney:1998}, yields an LSR bulk flow velocity
--17.0$\pm$4.6 \kms\ with upwind direction (\glong, \glat) =
(2.3\deeg, --5.2\deeg) and an LSR LIC vector of --15.7 \kms,
upwind (\glong, \glat) = (346.0\deeg, 0.1\deeg).} The flow
velocity is somewhat sensitive to the star sample because a
velocity gradient between the upwind and downwind directions
indicates the flow is decelerating \citep{FrischSlavin:2006}.

Data for the nearest cloudlets have been presented in a series of
studies by \cite{RLI,RLII,RLIII}.
Focusing only on UV observations of cloudlets within 15 pc as a
predictor of past and future variations in the Galactic
environment, we find a range of temperatures $T$ and turbulent
velocities $\xi$, $T$ = 1700--12,600 K and $\xi$ = 0--5.5 \kms,
with a mean temperature 6780$\pm$190 K, comparing favorably to the
LIC temperature 6300$\pm$340 K inferred by spacecraft
\citep{Witte04}. Only $\sim$35\% of space within 10 pc of the Sun
is filled with neutral gas if this material has the same density
as the LIC \citep[$\sim$0.20 \cc,][]{FrischSlavin:2003}, and if
\NDI/\NHI\ = $1.5 \times 10^{-5}$.  The mean cloud lengths are
0.9$\pm$0.3 pc. At a relative Sun-cloud velocity of 19 \kms, such
a distance is traversed in $\sim$47,000 yr.  The ISM filling
factor $\tilde{f}$ found locally varies from $\tilde{f} \sim 0.60$
towards $\alpha$ Aql (\glong, \glat, $d$ = 47.7\deeg, --8.9\deeg,
5.1 pc), to $\tilde{f} \sim 0.29$ in the opposite direction
towards Sirius (\glong, \glat, $d$ = 227.2\deeg, --8.9\deeg, 2.6
pc). Here, $\tilde{f}$ is the fraction of space filled with ISM if
all ISM has a density of 0.2 \cc.

Several sets of data indicate that the ISM within $\sim$3--10 pc
is not uniform.  The ratio \NFeII/\NDI\ varies between the
downwind and upwind LSR directions by up to a factor of $\sim$8,
apparently from ionization or abundance variations
\citep{Frisch:2004igpp}. The ISM temperature within 5 pc varies by
over a factor of 4 \citep{RLIII}.  If the cloud in front of the
nearest star $\alpha$ Cen also extends in front of $\alpha$ Oph,
as indicated by their common velocity, and is uniform, then \CaII\
and \HI\ data suggest a density $n > 5$ \cc\ \citep{Frisch:2003},
in contrast to the LIC density $\sim$0.3 \cc. Velocity variations
of 10 \kms\ or more are also found along several individual
sightlines \citep{FGW:2002,RLIII}.


A series of  radiative transfer models appropriate for the
radiation field and physical properties of the low column density
material close to the Sun show that equilibrium occurs for a range
of ionization levels in low density ISM
\citep{SlavinFrisch:2002,FrischSlavin:2006}. Furthermore, the
boundary conditions of the heliosphere will vary as it traverses
low density ISM strictly because of ionization variations within
the cloud. These models generally consider clouds with
\ntot$\ltsim$0.3 \cc, \NHI\ $< 10 ^{18}$ \cmtwo, a local radiation
field consistent with observations of the diffuse radiation field
between 3000 \AA\ and 0.5 keV at the solar location, and allow for
additional radiation emitted by a possibly magnetized conductive
interface between the local warm gas and adjacent very hot Local
Bubble plasma. These models predict ISM equilibrium conditions for
\nHI\ = 0.16--0.26 \cc, hydrogen ionization levels \chiH\ =
\HII/(\HI+\HII) = 0.19--0.34, and cloud temperatures of 4900--8300
K.

The time of the Sun's entry into, and exit from, the LIC can be
estimated using observations of \HI\ and \DI\ towards nearby stars
combined with \nHI\ derived from the Slavin \& Frisch radiative
transfer models. These models of the LIC indicate that \nHI\ =
0.19--0.21 \cc\ and \np\ $\sim 0.1$ \cc\ at the solar location,
and that \nHI\ decreases by $<$20\% between the Sun and surface of
the LIC. Assuming a constant LIC density of \nHI\ = 0.2 \cc\ and
using the limits on the LIC component towards 36 Oph \citep[\NHI\
$< 6 \times 10^{16}$ \cmtwo,][]{36Oph}, we infer that the distance
to the LIC surface in this direction, as defined by a velocity
discontinuity in the gas, is $<$0.1 pc, suggesting the Sun will
exit the LIC in less than 3700 yrs.

The entry of the Sun into the LIC can be calculated after
transforming into the LSR frame and assuming a LIC morphology
\citep[e.g.][]{Frisch94}. If the LIC velocity vector is
perpendicular to the surface, and assuming \NHI\ $=4.0 \times
10^{17}$ \cmtwo\ towards $\alpha$ CMa \citep{Hebrardetal:1999},
the Sun will have entered the LIC $\sim$6700/11,500 yr ago for the
Hipparcos/Standard solar apex motion, respectively. The assumed
column density requires additional ISM near the LIC velocity
towards the downwind stars $\alpha$ Aur and $\chi ^ 1$ Ori.
Alternatively, the LIC column densities towards downwind stars can
be used to define a plane that advances through space with the LIC
velocity vector. The useful downwind stars for this estimate are
$\alpha$ CMa, $\alpha$ CMi, $\chi ^ 1$ Ori, and $\alpha$ Aur,
where column densities for the LIC are, respectively, log \NHI =
17.60, 17.90, 17.80, 18.26 \cmtwo\ based on D/H $=1.5 \times
10^{-5}$ and data by \citet{Hebrardetal:1999} and \citet{RLII};
the selection of any three of these four stars, and assuming \nHI\
= 0.2 \cc, suggests that the Sun entered the LIC about
28,000--30,250 yr ago.
When uncertainties of $\sim$30\% are incorporated to reflect the
various assumptions, these estimates suggest that the Sun has
entered the interstellar cloud component at the LIC velocity
sometime within the past 40,000 yr,
and will exit it sometime within the next 4000 yr.

The star $\chi^1$ Ori is within 15\deeg\ of the downwind
direction, and shows a cloud with a relative Sun-cloud velocity of
21.6 \kms. The Sun and this cloud would have first crossed paths
$\sim$47,000 yr ago for \nHI\ = 0.2 \cc.  Beyond $\chi^1$ Ori, the
next neutral gas in the downwind direction is over 50 pc away.
Allowing for $\sim$30\% uncertainties and possible gaps between
clouds, the Sun would have entered the CLIC within the past
$\sim$60,000/$\tilde{f}$ yr.

\begin{figure}
   \includegraphics[angle=-90, width=0.5\textwidth]{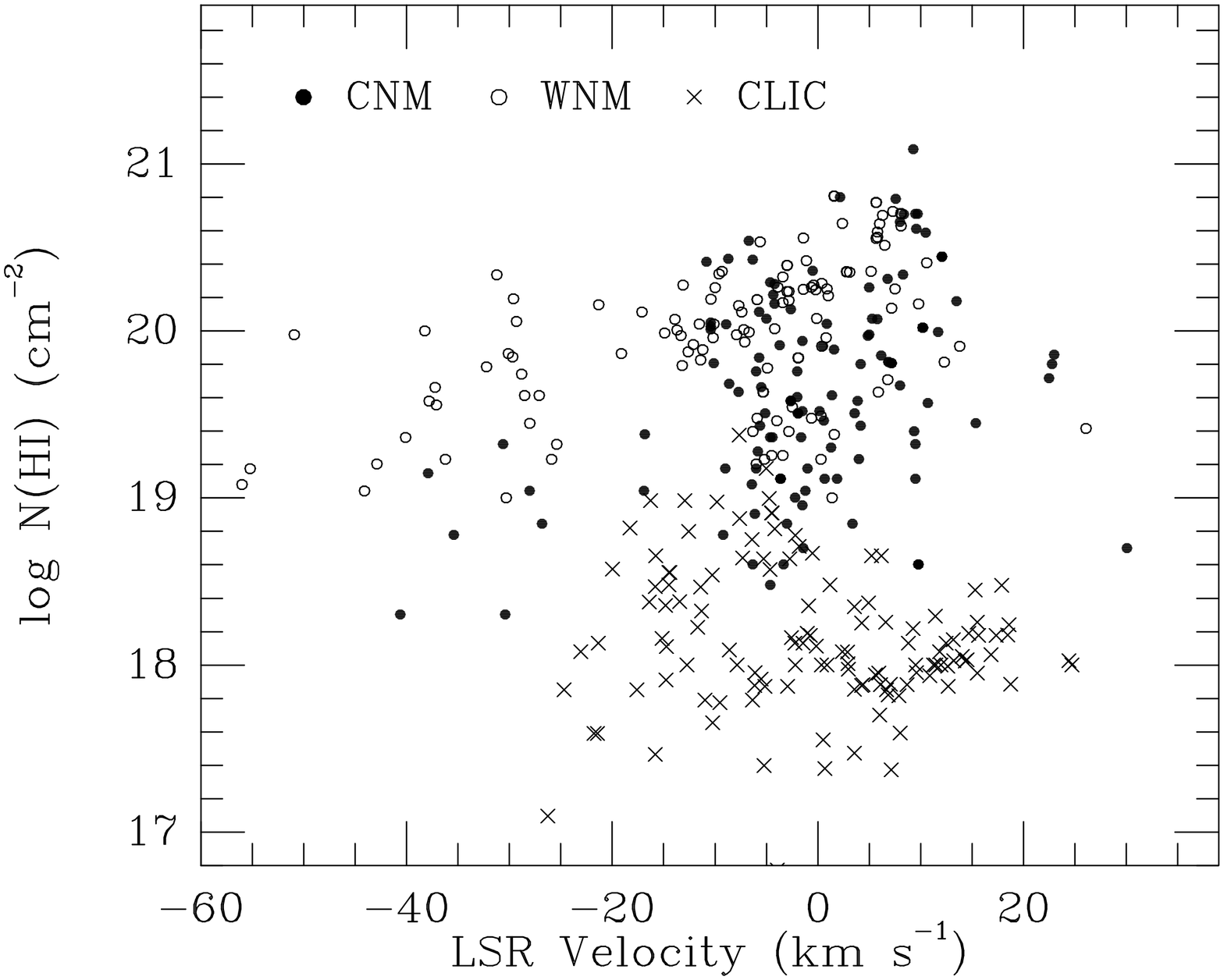}
   \caption{Velocity distribution of the CLIC versus CNM and WNM:
The CLIC components are plotted with the symbol ``$\times$''.  The
CLIC components shown here are towards stars within 50 pc. The LSR
velocities are derived from \HI, \DI, and optical \CaII\ data in
\citet{FGW:2002,RLI,RLII,Woodetal:2005} (Standard solar apex
motion used). Points based on \NDI\ (or \NCaII) data assume
\NDI/\NHI$ = 1.5 \times 10^{-5}$ (or \NCaII/\NHI$ = 1.0 \times
10^{-8}$). Filled and open circles show the CNM and WNM \HI\
components, respectively \citep[from][]{HTI}.  Only high latitude
HI sightlines are included ($|\mbox{b}|>$25\deeg). Note that the
CNM and WNM data sample may contain non-local components with
velocities affected by differential galactic rotation.  By
selecting the high latitude HI sightlines, the sample is more
likely restricted to nearby regions, where galactic rotation
effects make minimal contributions to the CNM and WNM component
velocities.} \label{fig:2}
\end{figure}

\subsection{Variations in the Global ISM \label{ss:globalism}}

An ISM with a wide range of properties is found within $\sim$350
pc of the Sun, including low density hot gas in the Local Bubble
that emits soft X-rays.  UV and radio observations of low column
density ISM show that a range of ISM types are possible at low
column densities. The ISM within that distance provides a model
for the types of ISM the Sun may encounter over timescales of
several million years. The Sun will move $\gtsim$16--20 pc through
the LSR per million years, and interstellar clouds (with
velocities of up to $\gtsim$100 \kms) may move hundreds of
parsecs.\footnote{A velocity of 1 \kms\ corresponds to $\sim$1
pc/Myrs.}

One example for ISM structure in this range is provided by
\citet{Welty99} who compare optical and UV ISM data for the star
23 Ori (300 pc), which is in the direction of the Orion-Eridanus
soft X-ray superbubble, and find a complex system of cloudlets
showing a wide range of properties, representing typical diffuse
ISM.  Twenty-one cloudlets with LSR velocities that range from
--120 to +8 \kms\ are found. Four low velocity clouds (positive
\Vlsr) at $\sim$100 K are present. They are massive (log \NH\ =
20.7 \cmtwo), moderately dense (10--15 \cc), suggesting a cloud
thickness of $\ltsim$15 pc, and primarily neutral. Crossing such a
cloud might take the Sun $\sim$1 Myrs.
Warmer dense clouds ($\sim$ 15--20 \cc, primarily neutral,
$\sim$3000 K) with thicknesses $\ltsim$1 pc are found at
moderately low velocities (\Vlsr\ $\sim -17 \rightarrow$ 0 \kms).
At higher velocities (\Vlsr\ $\sim -60 \rightarrow$ --17 \kms) low
density gas (\nH\ = 1--6 \cc) is found. This material appears to
be in thin sheets with thicknesses of 0.001--0.04 pc. Warm
($\sim$8000$\pm$2000 K), rapidly moving (\Vlsr\ $= -130
\rightarrow$ --100 \kms) shocked low column density clouds are
also seen.  This gas is partially ionized (\nel\ = \nH\ = 0.4--0.5
\cc) and arises in clouds with thicknesses of about 0.005--0.12
pc. The ionization of this high velocity gas suggests an
interstellar radiative shock where the gas is not in ionization
equilibrium (since the collisionally ionized species show $T\sim
25$,000 K, but Doppler $b$-values indicate 6000--12,000 K).

The Millennium Arecibo \HI\ 21-cm radio survey of warm neutral
material (WNM) and cold neutral material (CNM) also provides a
comparison sample for the CLIC \citep{HTI,HTII}. Figure
\ref{fig:2} shows the LSR velocities of CLIC components observed
in the optical and UV \citep[data
from][]{FGW:2002,RLII,RLIII,Woodetal:2005} compared to CNM and WNM
velocities. Except for extra high-latitude infalling ISM flows at
$v < -25$ \kms\ \citep{LockmanGehman:1991}, the kinematics of the
CLIC are similar to WNM and CNM.  If viewed from the outside, the
CLIC would appear as a medium velocity flow (17--20 \kms) with low
column densities (\NHI\ $< 10^{19}$ \cmtwo). The WNM has
upper-limit kinetic temperatures of 500 K to over 10,000 K, and
median column densities of $1.3 \times 10^{20}$ \cmtwo. The
dominance of low mass warm clouds at intermediate velocities
($<$--17 \kms) seen in Fig.\ \ref{fig:2}, combined with the fact
that $\sim$60\% of the \HI\ is WNM, suggest that warm low density
clouds are the most likely to be encountered by the Sun over the
next million years.

The recent discovery of cold ($<$100 K) tiny neutral cloudlets in
the ISM, \NHI\ $\sim 10^{18}$ \cmtwo, including one towards 3C286
at a velocity within 1.5 \kms\ of the G-cloud velocity
\citep{StanimirovicHeiles:2005}, shows that tiny cold neutral
clouds are widespread but infrequent. CNM components with \NHI\
$\ltsim 10^{18}$ \cmtwo\ and densities $\sim$20 \cc, similar to
values found by Welty et al.\ towards 23 Ori, would have
thicknesses of $<$0.02 pc and if at rest in the LSR would perturb
the heliosphere boundary conditions on timescales of $\sim$100 yr.

\begin{deluxetable}{ccccccc}
\tablecaption{Model Boundary Parameters\label{tbl-para}}
\tablecolumns{7} \tablewidth{0pt} \tablehead{
  \colhead{\#} &
  \colhead{$n_{{\rm H}^{\rm o}}$} & \colhead{$n_{{\rm H}^+}$} &
  \colhead{$n_{\rm tot}$} & \colhead{\chiH} &
  \colhead{$v_{\rm LISM}$} &  \colhead{$T_{\rm LISM}$} \\
  \colhead{} &
  \colhead{(cm$^{-3}$)} & \colhead{(cm$^{-3}$)} & \colhead{(cm$^{-3}$)} &
  \colhead{} &
  \colhead{(km s$^{-1}$)} & \colhead{(K)}  }
\startdata
\prlocbub & 0.00 & 0.005& 0.005& 1.00 &  13.4&1260000\\
\hdsub    & 0.14 & 0.10 & 0.24 & 0.42 &   8.3& 7000  \\
\prtwo    & 0.24 & 0.04 & 0.28 & 0.14 &  15  & 3000  \\
\preight  & 0.24 & 0.04 & 0.28 & 0.14 &  26  & 7000  \\
\prtwelve & 0.216& 0.047& 0.26 & 0.18 &  26  & 7000  \\
\prsixt   & 0.242& 0.074& 0.32 & 0.23 &  26  & 7000  \\
\preleven & 0.24 & 0.10 & 0.34 & 0.29 &  26  & 7000  \\
\prfift   & 0.235& 0.106& 0.34 & 0.31 &  26  & 7000  \\
\acen     & 0.14 & 0.10 & 0.24 & 0.42 &  25  & 5650  \\
\azz      & 0.14 & 0.10 & 0.24 & 0.42 &  26  & 8000  \\
\prone    & 0.04 & 0.04 & 0.08 & 0.50 &  15  & 3000  \\
\prseven  & 0.04 & 0.04 & 0.08 & 0.50 &  26  & 7000  \\
\azzw     & 0.10 & 0.10 & 0.20 & 0.50 &  26  & 8000  \\
\hivel    & 0.40 & 0.40 & 0.80 & 0.50 & 100  & 8000  \\
\hidensa  &11.00 & 0.15 &11.15 & 0.01 &  26  &  100  \\
\hidensc  &15.00 & 0.20 &15.20 & 0.01 &  26  &   10  \\
\hidensb  &15.00 & 0.20 &15.20 & 0.01 &  26  & 3000  \\
\xiboo    & 0.14 & 0.10 & 0.24 & 0.42 &  31.5& 5650  \\
\oph      & 0.14 & 0.10 & 0.24 & 0.42 &  37.7& 5650  \\
\evlac    & 0.14 & 0.10 & 0.24 & 0.42 &  45.2& 7000  \\
\vir      & 0.14 & 0.10 & 0.24 & 0.42 &  50.8& 7000  \\
\eind\tablenotemark{a}
          & 0.14 & 0.10 & 0.24 & 0.42 &  68  & 8000  \\
\cyg      & 0.14 & 0.10 & 0.24 & 0.42 &  86  & 8000  \\
\colda    & 0.24 & 0.04 & 0.28 & 0.14 &  50.8&   10  \\
\coldb    & 0.96 & 0.04 & 1.00 & 0.04 &  50.8&   10  \\
\enddata
\tablenotetext{a}{Model \eindmod, $T_{\rm SW}$= $2 \times 10^5$ K.
Model \eindmodm, $v_{\rm SW}$= 500 \kms.}
\end{deluxetable}

\subsection{Modeled Clouds\label{ss:models}}

The above discussion of the ISM in the solar neighborhood, and
close to the Sun, provides the basis for selecting a
representative set of boundary conditions for modeling. The
heliosphere configuration has been modeled for \nummodels\ cloud
types with densities varying from 0.005--15 \cc, ionizations
ranging up to 100\%, and relative Sun-cloud velocities of up to
100 \kms. The boundary parameters are listed in Table
\ref{tbl-para}. Most of the assumed cloud types are warm and low
density clouds, but the possible velocities vary by an order of
magnitude.  The Sun moves through the LSR at $\sim$14--19.5 \kms,
while warm diffuse clouds have LSR velocities 20--60 \kms, and
higher.  In particular, the cloud cluster near the Sun shows
depletions characteristic of shocked interstellar gas
\citep{Frisch:1981,Frisch:2004igpp}, where large peculiar motions
might be expected.  Hence relative Sun-cloud velocities may range
over an order of magnitude, leading to appreciable variations in
the heliosphere morphology.

Table \ref{tbl-para} gives the LISM boundary conditions as neutral
hydrogen number density \nH, proton number density \np, and
heliocentric velocity $v_{\rm LISM}$ and temperature $T_{\rm
LISM}$ of the interstellar wind.  \ntot\ is the total hydrogen
density, and \chiH\ = \np/\ntot\ the interstellar hydrogen
ionization fraction.
Model \prlocbub\ (\ntot=0.005 \cc, \chiH=1, $v$=13.4 \kms, log
$T$=6.1 K) represents the hot plasma interior of the Local Bubble
\citep{Snowdenetal:1997}. Model \hdsub\ also tests subsonic
interstellar conditions, but with a 42\% partially ionized dense,
warm ISM (\ntot=0.24 \cc, $T$=7000 K). The velocity in model
\hdsub\ (8.3 \kms) is comparable to the heliocentric velocity of
the blue-shifted cloud found towards $\epsilon$ CMa and $\alpha$
CMa \citep{GryJenkins:2001}, which may have been the first warm
cloud encountered by the Sun as it exited the plasma interior of
the Local Bubble.

Models \prtwo--\azz\ test the effect of small density
(\ntot=0.24--0.34 \cc), warm temperature (3000--8000 K), and
velocity (15--26 \kms) variations on the heliosphere morphology,
with varying ionizations (\chiH=0.14--0.42), representing the warm
low density ISM described earlier. Neutral clouds with $T \sim
3000$ K are widespread and evidently thermally unstable in the
absence of magnetic pressure \citep{Heiles01}. Models
\preight--\azz, with $v \sim 26$ \kms, represent variants of the
contemporary heliosphere whose interstellar boundary densities
fall within the constraints of the observations. Model \acen\
corresponds to the $\alpha$ Cen environment
\citep{LinskyWood96,Woodacen}, provided the ISM spreads uniformly
throughout the sightline towards this nearest star. Model \azz\ is
based on the local cloud towards the white dwarf star REJ 1032+532
(Holberg et al. 1999).

Models \prone--\hivel\ represent the effect on the heliosphere of
density (\ntot=0.08--0.8 \cc) and velocity (15--100 \kms)
variations in 50\% ionized interstellar hydrogen at warm
temperatures (3000--8000 K), such as might be expected for
kinematically perturbed low density gas. Models
\hidensa--\hidensb\ test the expected dramatic differences in
heliosphere configuration anticipated from the encounter with a
denser (\ntot=11--15 \cc) neutral (\chiH=0.01) cold or tepid
(T=10--3000 K) interstellar cloud at the LIC velocity.  Model
\hidensa\ is based on the strong line, low velocity gas towards 23
Ori \citep{Welty99}, while model \hidensb\ is based on both the
warm low velocity gas towards 23 Ori, and the thermally unstable
warm \HI\ gas observed at 21 cm (Heiles 2001). Model \hidensc\ was
calculated solely for comparison with model \hidensb, representing
a temperature reduction by a factor of 300 without a corresponding
density increase that typical Galactic values of the roughly
constant product $n_{tot} \, T_{LISM}$ would suggest.

Models \xiboo--\cyg, together with \hdsub, test the parameter
space around a partially ionized, diffuse interstellar cloud
(\chiH=0.42; \ntot=0.24 \cc), sampling velocity variations from 8
to 100 \kms. Changes that arise from small variations in the solar
wind are also considered (models \eindmod\ and \eindmodm). Model
\hivel\ is an example of cooled high velocity shocked gas such as
a superbubble shell formed from a supernova shock sweeping up the
ISM in a starforming region. Its parameters are based on high
velocity gas observed towards 23 Ori and $\zeta$ Ori
\citep{Welty99,Welty02}. Finally, models \colda--\coldb\ test a
cold (10 K), high velocity (51 \kms) regime, again at lower values
of $n_{tot} \, T_{LISM}$ than typically encountered, in order to
discuss low-temperature regimes without getting into the
complications of very high density heliospheres.


\section{INDIVIDUAL MODEL RESULTS\label{ss:exmodels}}

To characterize the large-scale heliospheres that result when the
Sun is embedded in the different parts of the ISM as described
above, we make use of the multi-fluid model developed by
\citet{Zank96}. The multi-fluid code simultaneously solves the
time evolution of four interpenetrating fluids. One fluid
represents the protons of the interstellar plasma component as
well as the solar wind plasma. The remaining fluids model three
thermodynamically distinct populations of neutral H. Each of the
neutral fluids interacts with the plasma through resonant \ce,
using the \citet{Fite} cross section, and all neutrals are
subjected to photoionization which depends on the squared distance
to the Sun. Radiation pressure is assumed to balance gravity. For
a detailed description of the numerical model, and the underlying
physics, see \citet{Zank96} and \cite{Zank99}.

Modeling of heliospheric neutrals in the multi-fluid code as a
superposition of three independent neutral fluids is an
approximation to the general, non-Maxwellian neutral distribution
function. Neutrals can also be calculated on a kinetic level
without such an approximation
\citep[e.g.,][]{Baranov1993,Mueller00}. Recent comparison studies
\citep{Alexashov05,Heerikhuisen06} have shown that the multi-fluid
and kinetic methods are in essential agreement in most of the fine
details concerning heliospheric geometry and shock locations, and
even the distributions look similar. For this reason, we adhere to
the computationally less costly multi-fluid method for this study.

For the numerical models, the solar wind at 1 AU is assumed to be
independent of longitude, latitude, and time, with values of 5.0
\cc\ for the plasma density, a temperature of $10^5$ K, and a
radial velocity of 400 \kms. There are two models (\eindmod\ and
\eindmodm) with slight variations on these solar wind conditions
($2 \times 10^5$ K and 500 \kms, respectively). The models of the
interaction of the solar wind with the ISM are carried out in a
heliocentric frame of reference, and are effectively
two-dimensional as we assume azimuthal symmetry about the
stagnation axis (the axis parallel to the LISM flow that contains
the Sun). To satisfy this assumption we also neglect heliospheric
and interstellar magnetic fields. The interstellar medium is
prescribed as four boundary conditions at a suitably large
distance from the Sun (a typical value is 1000 AU for heliospheres
where the interstellar bow shock is less than 500 AU from the
Sun). The boundary parameters are the LISM \HI\ and \HII\ number
densities, and the (common) hydrogen velocity and temperature.

\begin{figure}
\epsscale{1.00} \plotone{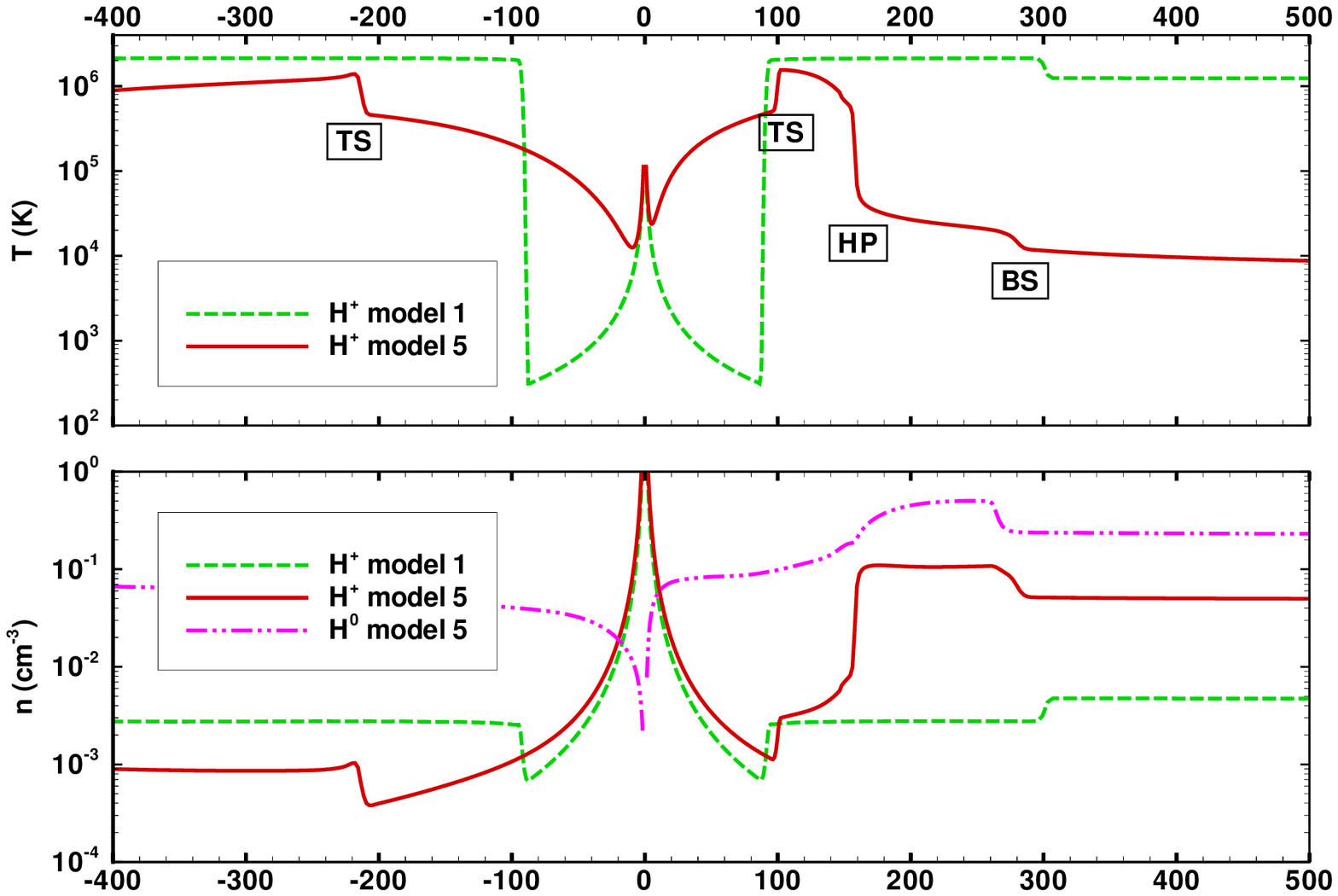} \caption{One-dimensional profiles
along the stagnation axis, with the Sun at center and the LISM
coming from right. Top: plasma temperature of model \prlocbub\
(the Local Bubble case; dashed) and model \prtwelve\ (representing
contemporary conditions; solid). The heliospheric boundaries of
model \prtwelve\ are marked in the plot. The bottom panel contains
the corresponding densities (plasma model 1, dashed; plasma model
5, solid; model 5 neutral H with dash-dot
pattern).\label{fig-1d15}}
\end{figure}

\subsection{Contemporary LISM\label{ss:contemporary}}

Inferred values of the contemporary interstellar boundary
parameters are $v = 26.3$ \kms\ and $T \sim 6300 \pm 340$ K
\citep{Witte96,Witte04}. The contemporary interstellar proton and
neutral H densities are not well constrained but should lie in the
range from 0.04--0.14 \cc\ and 0.14--0.24 \cc, respectively
\citep[e.g.][]{Zank99,SlavinFrisch:2002}, and models
\preight--\azz\ (Table \ref{tbl-para}) fit within these
contemporary constraints. We choose model \prtwelve\ (\np\ = 0.047
\cc, \nH\ = 0.216 \cc, $v$ = 26 \kms, and $T$ = 7000 K) as the
highlighted example representing the contemporary heliosphere.
This model has been described in some detail previously by
\citet{Mueller04}, whose Figure 1 displays two-dimensional maps of
plasma temperature and neutral density, featuring this model's
heliospheric boundaries together with the neutral hydrogen wall.
Here, we do not repeat this figure, but rather display the plasma
temperature (top panel) and density (bottom panel) along the
stagnation axis as solid lines in Figure \ref{fig-1d15}, together
with the neutral H density (dash-dotted line).

The heliospheric boundaries are clearly visible as discontinuities
in Figure \ref{fig-1d15}, and labels are provided next to the
temperature profile. The solar wind is supersonic at 1 AU, and
expands radially before undergoing a transition (temperature and
density increase) at the termination shock (TS). The TS is
asymmetric, with a nose distance of 99 AU and a tail distance of
216 AU. In the heliosheath, the region of the shocked and heated
solar wind, the solar wind plasma gets directed tailward, and is
separated from the interstellar plasma by the heliopause (HP),
with discontinuous density and temperature. The stagnation point
(the nose of the heliopause) is at 148 AU. The LISM is supersonic,
and consequently there is an interstellar bow shock (BS) at 285 AU
upwind.

The thermodynamically distinct plasma regions define the
characteristics of the three neutral fluids used in the four-fluid
model. The component 1 neutral population consists of neutrals
from the ISM. They typically are warm and of moderate bulk speed.
Neutrals born through \ce\ in the hot heliosheath between the TS
and the HP form a different neutral population (component 2) which
is hot, with correspondingly high thermal speeds. The third
component consists of neutrals born inside the TS in the
supersonic solar wind; correspondingly, component 3 neutrals are
fast and warm. While the approximation of the real neutral
distribution function throughout the heliosphere by a
superposition of three Maxwellians (the three neutral fluids)
introduces inaccuracies into the models, such a hydrodynamic
treatment represents the overall neutral H distribution well, as
evidenced for example by the successful matching of modeled and
observed \lya\ absorption by heliospheric \HI\ \citep{Wood00}.

Figure \ref{fig-1d15} also shows the temperature and density
profile of a plasma-only model (model \prlocbub, dashed lines, TS
at 90 AU, HP at 300 AU), which will be described in detail in the
next section. The contrast between the two plasma temperature
profiles is due to the effect of \ce: The pickup process (here,
the \ce\ of a solar wind proton with an interstellar neutral H
atom in the supersonic solar wind region) deposits energy into the
supersonic solar wind and reverses the effect of adiabatic cooling
(and also slows the solar wind), such that the effective solar
wind temperature of model \prtwelve\ turns upward, whereas model
\prlocbub\ follows an adiabatic cooling law. In both inner and
outer heliosheath, heat transport by neutral hydrogen with
subsequent secondary \ce\ introduces gradual temperature gradients
between the discontinuities, which are absent in model \prlocbub\
which has no neutrals, no \ce, and no anomalous heat transport
across the HP.

In the contemporary heliosphere (model \prtwelve), there is a
hydrogen wall between the BS and the HP, with a peak density of
0.502 \cc\ = 2.3 \nH\ (Figure \ref{fig-1d15} bottom, dash dot
line). The extra wall material consists of slower neutral hydrogen
born from \ce\ with the subsonic interstellar plasma downstream of
the BS (the outer heliosheath). The outer heliosheath plasma is
interstellar plasma that is slowed and heated at the BS, and
further affected by additional momentum loss and energy gain
through \ce\ of component 2 neutrals that cross from the inner
into the outer heliosheath. The hydrogen wall is accompanied by an
elevated plasma density (Figure \ref{fig-1d15} bottom, solid line)
because of the plasma slowdown.

The neutral atom density at the upwind TS is 0.098 \cc\ = 0.46
\nH, the latter ratio being called the filtration factor because
it links the interstellar density to the neutral density in the
inner heliosphere, accounting for the loss processes along its
path. The neutral density at 5 AU on the upwind stagnation axis is
0.036 \cc\ = 0.17 \nH. Closer to the Sun, photoionization and
solar wind charge exchange deplete neutral H exponentially. In the
tail direction, the stagnation axis re-populates slowly with
off-axis neutral H.

Models \preight--\azz\ of Table \ref{tbl-para} loosely fit within
the constraints of contemporary observations, and generally
represent ionization and density levels appropriate for low column
density ISM where both radiation field and abundances of cooling
trace elements might vary.  Table \ref{tbl-res} gives the
corresponding results for all models, and it can be seen that the
results from models \preight--\azz\ only vary modestly. The size
of the heliosphere changes, with HP locations ranging from
100--150 AU, and neutral hydrogen filtration varies from 0.3--0.5,
while the relative peak wall density remains essentially
unchanged.

\subsection{Hot Local Bubble\label{ss:locbub}}

The interior of the Local Bubble void is hot and nearly completely
ionized. The absence of interstellar neutral hydrogen in model
\prlocbub\ simplifies the heliospheric physics considerably,
because in all other models the atom-ion process of \ce\ generates
distinct features in the atom distribution function, and also
alters the plasma by coupling ions and atoms.

We adopt the ISM parameters of \np\ = 0.005 \cc, \nH\ = 0, $v$ =
13.4 \kms, and log T(K) = 6.1 (model \prlocbub\ in Table
\ref{tbl-para}), with the velocity based on the
\citet{DehnenBinney:1998} solar apex motion since the plasma is
assumed at rest in the LSR.  The speed of sound in such a plasma
is 190 \kms, and the Sun therefore moves subsonically through this
medium (Mach 0.07). In this case, the isotropic thermal
interstellar pressure dominates the ram pressure, and the
termination shock is spherical at a distance of 90 AU from the
Sun. This distance is comparable to that of the contemporary
heliosphere. The distance to the nose of the heliopause is 300 AU,
which makes this sheath very large in comparison to the
contemporary heliosphere above. The temperature in the sheath
reaches values as high as $2.2\times 10^6$ K. Figure
\ref{fig-1d15} contains the plasma temperature profile (top) along
the stagnation axis, and the plasma density (bottom), as dashed
lines. In the termination shock transition, the density jumps by a
factor of 3.8, and the wind speed decreases to 100 \kms.

In this model there are no neutral atoms in the entire system of
solar wind and interstellar medium, and hence the particle content
is different from that observed today. There are no pickup ions
(PUI) produced by \ce, and therefore there are no anomalous cosmic
rays, and no slowdown or heating in the supersonic solar wind
beyond the inner solar system. However, thanks to its large ram
pressure the solar wind plasma still provides an effective shield
of the solar system against the million-degree plasma of the Local
Bubble.

As a plasma-only model with subsonic interstellar boundary
conditions, model \prlocbub\ is ideally suited for comparison to
previous, analytical studies of the heliosphere under the
assumption of incompressibility of the flow outside of the TS. For
the simple case of a flow around a rigid sphere (as a stand-in for
the heliopause), the interstellar velocity on the stagnation axis
behaves as $v_{\infty}$ $(1-r_{\rm HP}^3$/$r^3$), where $r_{\rm
HP}$ is the distance to the HP nose, and $v_{\infty}$ the
uncontaminated interstellar flow velocity. This analytical
behavior matches the one of model \prlocbub\ very well (Figure not
shown here). In a detailed treatment, \citet{SuessNerney90}
calculate the flow streamlines from a realistic, pressure-balanced
TS outward. Their TS distance formula (equation (\ref{eq-TS2})
below) predicts it to be at 81.8 AU, which compares well with the
90 AU found in model \prlocbub, given that the latter (numerical)
model does not presuppose incompressibility.

It has to be noted that the models do not account for the
interplanetary magnetic field, as doing so requires a
three-dimensional treatment of the problem that is outside the
scope of this paper. In particular for model \prlocbub\ in which
the mitigating aspects of the neutral-plasma interaction are
absent, the pile-up of magnetic field at the nose of the HP
\citep[Axford-Cranfill effect; e.g.][]{NerneySuess93} may be a
contributor to the overall pressure balance in the upwind
directions of the inner heliosheath. The plasma would not
decelerate as quickly as 1/$r^2$ downstream of the TS, as is
approximately the case in model \prlocbub. Consequently, the
heliopause would be expected to shift further away than the
location identified in this paper. The unknown interstellar
magnetic field strength in the Local Bubble has the potential to
shift the pressure balance of the heliosphere as well.

\begin{figure}
\epsscale{1.00} \plotone{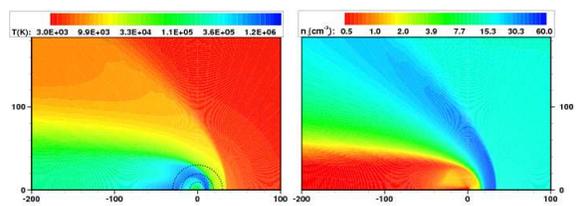} \caption{Two-dimensional maps
of plasma temperature (left) and neutral H density (right) for the
high density case \hidensb, with the Sun at center and the LISM
coming from right. The transition from white to medium gray in the
plasma temperature is the interstellar bow shock; the dark shades
are the hot heliosheath and heliotail. The orbits of Saturn,
Uranus, and Neptune are sketched as dotted lines.\label{fig-2d20}}
\end{figure}

\begin{figure}
\epsscale{1.0} \plotone{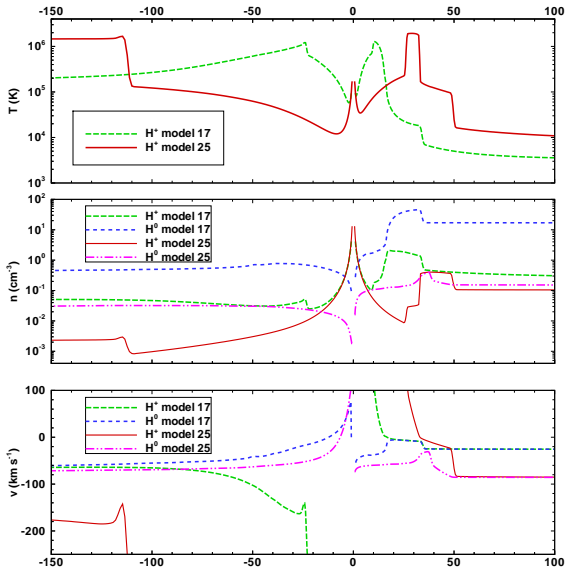} \caption{Similar to Figure
\ref{fig-1d15}, 1-D profiles of plasma temperatures (top), number
densities (middle), and parallel velocities (bottom), for model
\hidensb\ (high density case; long and short dashed), and model
\cyg\ (high velocity case; solid and dash-dot).\label{fig-1d1820}}
\end{figure}

\begin{figure}
\epsscale{1.00} \plotone{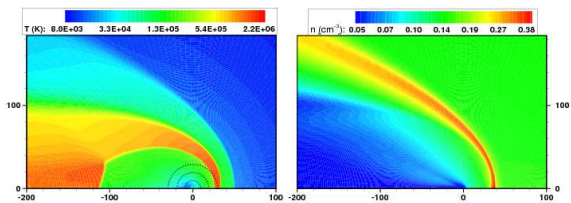} \caption{Two-dimensional maps
of plasma temperature (left) and neutral H density (right) for the
high speed case \cyg. Note the triple point at about (-105, 30).
The orbits of Saturn, Uranus, and Neptune are sketched as dotted
lines. In the neutral density (right panel), the hydrogen wall is
clearly visible.\label{fig-2dcyg}}
\end{figure}

\subsection{Dense ISM\label{ss:hidens}}

Three models in Table \ref{tbl-para} represent the effect of an
ISM that is denser than that of the contemporary heliosphere by a
factor of $\sim$50 (models \hidensa--\hidensb), but not as dense
as models investigated by \citet{Yeghikyan03} with $n_{\rm tot}$ =
100 \cc. Additionally, models \hivel\ and \coldb\ are denser by a
factor of 3--4 above the contemporary value. The high density
leads to a high interstellar ram pressure, and therefore the
resulting heliosphere tends to be smaller.

As an example, Figure \ref{fig-2d20} displays 2-D maps of the
hydrogen density and plasma temperature of model \hidensb\ (\nH =
15 \cc), and Figure \ref{fig-1d1820} shows stagnation axis
temperature, density, and velocity profiles (dashed). The TS is
asymmetric with a nose distance of 9.8 AU and a tail distance of
23 AU (Table \ref{tbl-res}). The HP is located at 16 AU, and the
BS at 34 AU upwind. The TS is weak, with an upwind compression
ratio of 1.8. Atypically, the HP is not a sharp temperature
gradient as in most other models, but the temperature profile is
more washed out by frequent \ce. The bow shock has a moderate
compression ratio of 3.

In spite of the small heliosphere, neutrals and protons are
coupled tightly because of the high neutral density, so that the
hydrogen wall starts immediately downstream of the BS. It has a
peak density of 3.1 \nH, but \ce\ is frequent enough that the
neutral density at the TS (the filtration factor) is 0.12 \nH,
which is among the lowest of the \nummodels\ models considered
here. Even so, the absolute neutral density is quite high, leading
to a pronounced solar wind slowdown. On the upwind stagnation
axis, the solar wind speed decreases to 260 \kms\ upstream of the
TS. In the tail region, the solar wind plasma slows to 160 \kms,
but frequent \ce\ decreases that value in the heliotail, to a
common plasma/neutral speed of $\sim$60 \kms\ by $\sim$100 AU
downwind of the Sun.

\begin{deluxetable}{cccccccccc}
\tablecaption{Model Results\label{tbl-res}} \tablecolumns{10}
\tablewidth{0pt} \tablehead{
  \colhead{\#} &
  \colhead{TS} & \colhead{HP} & \colhead{BS} & \colhead{TS$_d$} &
  \colhead{$f_{\rm peak}$} &  \colhead{$f_{\rm TS}$} &
  \colhead{$f_{\rm 5AU}$}  \\
  \colhead{} &
  \colhead{AU} & \colhead{AU} & \colhead{AU} & \colhead{AU} &
  \colhead{} & \colhead{} & \colhead{} }
\startdata
\prlocbub &  90 & 300 &     &  90 &     &      &       \\
\hdsub    & 259 & 402 &     & 371 & 1.1 & 0.12 & 0.04  \\
\prtwo    & 149 & 233 & 535 & 286 & 2.1 & 0.29 & 0.10  \\
\preight  &  85 & 132 & 225 & 191 & 2.3 & 0.54 & 0.23  \\
\prtwelve &  99 & 148 & 285 & 216 & 2.3 & 0.46 & 0.17  \\
\prsixt   &  74 & 115 & 198 & 164 & 2.3 & 0.40 & 0.20  \\
\preleven &  83 & 110 & 250 & 178 & 2.4 & 0.38 & 0.13  \\
\prfift   &  69 & 100 & 186 & 152 & 2.4 & 0.37 & 0.16  \\
\acen     &  99 & 137 & 280 & 207 & 2.5 & 0.29 & 0.10  \\
\azz      &  79 & 104 & 230 & 166 & 2.3 & 0.40 & 0.17  \\
\prone    & 253 & 358 & 812 & 463 & 2.1 & 0.28 & 0.08  \\
\prseven  & 144 & 197 & 365 & 331 & 2.4 & 0.50 & 0.18  \\
\azzw     &  86 & 119 & 242 & 180 & 2.3 & 0.39 & 0.17  \\
\hivel    &  11 &  14 &  21 &  52 & 3.3 & 0.82 & 0.60  \\
\hidensa  &  14 &  26 & 100 &  37 & 3.0 & 0.12 & 0.08  \\
\hidensc  & 8.2 &  12 &  23 &  28 & 7.0 & 0.14 & 0.10  \\
\hidensb  & 9.8 &  16 &  34 &  23 & 3.1 & 0.12 & 0.09  \\
\xiboo    &  91 & 126 & 227 & 227 & 2.9 & 0.35 & 0.12  \\
\oph      &  62 &  85 & 136 & 173 & 3.2 & 0.50 & 0.21  \\
\evlac    &  52 &  69 & 108 & 159 & 3.3 & 0.59 & 0.26  \\
\vir      &  45 &  62 &  92 & 152 & 3.5 & 0.68 & 0.31  \\
\eind     &  32 &  44 &  63 & 122 & 2.8 & 0.97 & 0.51  \\
\eindmod  &  32 &  44 &  63 & 121 & 2.9 & 0.96 & 0.52  \\
\eindmodm &  39 &  54 &  79 & 139 & 3.0 & 0.82 & 0.42  \\
\cyg      &  26 &  34 &  50 & 112 & 2.7 & 1.00 & 0.64  \\
\colda    &  38 &  44 &  77 & 141 & 5.3 & 0.87 & 0.25  \\
\coldb    &  21 &  31 &  46 &  72 & 3.4 & 0.88 & 0.54  \\
\enddata
\end{deluxetable}

\subsection{High Velocity ISM\label{ss:hivel}}

Several models in Table \ref{tbl-para} test the response of the
heliosphere on which a high velocity ISM impinges (models \hivel\
and \xiboo--\coldb). The high velocity generates a large ram
pressure, making the resulting heliosphere smaller, similar to,
but more elongated than, the high density cases. We present model
\cyg\ with $v$ = 86 \kms\ in more detail here. If the
corresponding cloud had a thickness comparable to the high
velocity ISM towards 23 Ori ($<$0.12 pc, \S \ref{ss:globalism}),
it would pass over the Sun in less than $\sim$1400 yr. Figure
\ref{fig-2dcyg} displays 2-D maps of hydrogen density and plasma
temperature of model \cyg, showing the heliospheric boundaries and
features clearly. The 1-D stagnation axis profiles in temperature,
density, and velocity are displayed in Figure \ref{fig-1d1820}.

In most models, the TS, taken as a 3-D surface, is spherical
(model \prlocbub), or nearly spherical with an upwind/downwind
asymmetry. These cases are characterized by a heliosheath and
heliotail plasma that are subsonic throughout. In contrast to
this, the shape of the TS of a high velocity ISM heliosphere, such
as model \cyg, is qualitatively different, now resembling a rocket
shape. The initially subsonic plasma at the nose of the
heliosheath accelerates to supersonic speeds in the nozzle-shaped
region between the TS and the HP. However, to match the subsonic
heliotail plasma and the supersonic heliosheath plasma requires
both a shock to decelerate the flow and a tangential discontinuity
to adjust the density. A characteristic triple point occurs where
heliosheath shock, termination shock, and the tangential
discontinuity meet. Figure \ref{fig-2dcyg} shows this morphology
in the example of model \cyg.

The TS is highly asymmetric, with a nose distance of 26 AU and a
tail distance of 112 AU (Table \ref{tbl-res}). The upwind TS
compression ratio $s$ is $s=2.9$. The HP is at 34 AU, and the BS
at 50 AU upwind. The bow shock is quite strong, with a post shock
plasma speed of 24 \kms\ (Figure \ref{fig-1d1820}, bottom panel),
temperature of $10^5$ K (top), and a compression ratio of 3.4
(middle). Because of the large neutral velocity in the post-bow
shock region, the neutral mean free path (mfp) for charge exchange
is initially $\sim$30 AU, larger than the outer heliosheath, and
shortens only gradually as the effective neutral velocity
decreases to 31 \kms. Consequently, the model \cyg\ hydrogen wall
between the BS and the HP is not very thick, but reaches a peak
density of 2.7 \nH\ about $\sim$11 AU downstream of the BS. This
distance from the BS is of the same magnitude as in the more
moderate cases, however, in the high speed case, it brings the
peak close to the HP already.
As the TS is so close to the hydrogen wall, the filtration factor
is 1.0, i.e.\ the neutral density at the TS equals that of the
LISM (and is still 0.6 \nH\ at 5 AU, Figure \ref{fig-1d1820},
middle panel). These filtration factors close to unity (models
\hivel, \eind\ -- \coldb\ in Table \ref{tbl-res}) seem only
possible when a high interstellar velocity combines with a modest
or low density so that the peak hydrogen wall occurs close to the
HP without room for depletion of neutral H between peak and HP. In
the similar sized dense heliosphere (\S \ref{ss:hidens}), the \ce\
mfps are shorter, the peak hydrogen wall is attained farther away
from the HP, and \ce\ upwind close to the HP spreads the H flow
and leads to a density decrease already before the \HI\ flow
crosses the HP (Figure \ref{fig-1d1820}).

Finally, we note that for the high speed models \hivel\ and \eind\
-- \coldb, the original numerical grid with a 5\deeg\ angular
resolution is too coarse, leading to errors during the transport
along directions with high velocity components. In these cases, we
choose a grid with a 2\deeg\ angular resolution, which cures the
problem so that pre-BS density values on the stagnation axis are
within 10\% or better of the LISM values. They are always more
accurate in off-axis directions.

\section{MODEL CORRELATIONS \label{ss:results}}

In addition to the individual model results discussed in the
preceding section, Table \ref{tbl-res} contains the same key
results characterizing the boundary locations and neutral H
content for all \nummodels\ models. The model numbers refer to the
corresponding boundary parameters listed in Table \ref{tbl-para}.
Model \prlocbub\ stands out from the rest in that neutrals are
absent. For the subsonic models \prlocbub\ and \hdsub\ the LISM
pressure is dominated by the thermal pressure. All other models
are ram-pressure dominated. This is among the reasons why the
results discussed below cover only a subset of the vast parameter
space.

\subsection{Plasma structure}

When comparing the \nummodels\ models, an obvious result is the
variation in the size of the heliosphere, as expressed in the
location of upwind TS, HP, and BS in Table \ref{tbl-res}, as well
as the distance of the downwind termination shock (TS$_d$). These
distances are set by balancing the solar wind pressure and
interstellar pressure \citep[e.g.][]{Holzer:1989}. A small
heliosphere is caused by a large LISM pressure. Models \hdsub,
\prtwo, \prone\ and \prseven\ are especially large due to a lower
LISM plasma ram pressure, namely, a low LISM velocity in the case
of models \hdsub, \prtwo\ and \prone, and a low density for models
\prone\ and \prseven. Overall, the upstream distances from the Sun
to the TS range from 8 to 260 AU, and those to the HP range from
12 to 400 AU (see Table \ref{tbl-res}). The bow shock of a cool,
slow, tenuous LISM (model \prone) is as far as 810 AU away from
the Sun.

\begin{figure}
\epsscale{1.0} \plotone{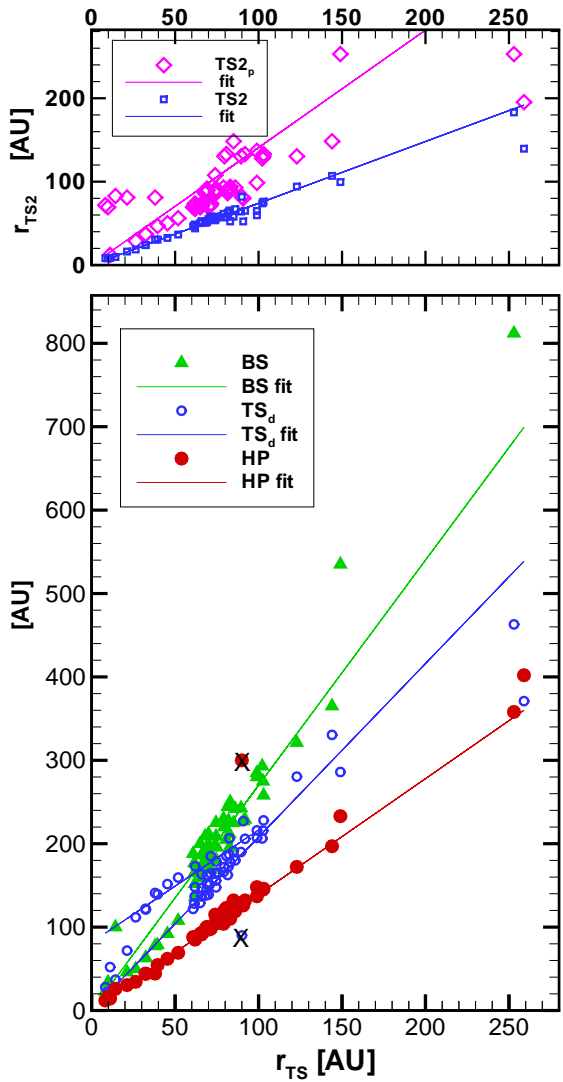} \caption{Correlation between the
upwind termination shock location $r_{TS}$ and other distances,
including, in the bottom panel, the upwind heliopause HP (red
circles), the upwind bow shock (green triangles), and the tailward
TS distance (open blue circles), which are all model results.
Model 1 results are marked with ``$\times$''. The upper panel
shows the TS correlations with the distance $r_{TS2}$
theoretically derived from plasma pressure balance (pink diamonds)
and from total pressure balance (blue squares), together with
empirical linear fits. Note the aspect ratio of 2:1 in both
panels.\label{fig-tscorr}}
\end{figure}

Because the overall system is pressure balanced, the locations
(heliocentric distances) of TS, HP, and BS are correlated with one
another. Taking all models except model \prlocbub, and adding the
results of another systematic (as yet incomplete) parameter study
with $v$ = 26.24 \kms, the correlation between $r_{\rm TS}$, the
distance of the upwind TS, with the distance of the upwind
heliopause $r_{\rm HP}$, is
\begin{equation}\label{eq-hpts}
r_{\rm HP} = (1.39\pm 0.01) \,\, r_{\rm TS},
\end{equation}
obtained with a linear regression analysis after ascribing
uncertainties to $r_{\rm TS}$ and $r_{\rm HP}$ due to grid
resolution and HP stability. The intercept value of the analysis
is consistent with zero. The omitted model \prlocbub\ ratio
$r_{\rm HP}/r_{\rm TS} = 3.3$; the other subsonic case \hdsub\ is
an outlier with a ratio of 1.55. The filled circles in Figure
\ref{fig-tscorr} represent each model as a (TS, HP) pair, and the
corresponding straight line is the linear fit of equation
(\ref{eq-hpts}). Model \prlocbub\ is additionally marked with an
``$\times$''.

For models with a supersonic LISM, the upwind bow shock $r_{\rm
BS}$ is located about twice as far as the heliopause,
\begin{equation}\label{eq-bshp}
r_{\rm BS} = (1.95\pm 0.05) \,\, r_{\rm HP}.
\end{equation}
A direct correlation between TS and BS yields $r_{\rm BS} =
(2.70\pm 0.09) \, r_{\rm TS}$. In deriving the BS correlations,
models \prtwo\ and \prone\ were excluded as the largest models;
the above relation underpredicts the BS distance of model \prtwo\
and \prone\ by 81 AU and 114 AU, respectively. Figure
\ref{fig-tscorr} shows the model (TS, BS) pairs as triangles. In
contrast to the excellent HP data fit, the BS locations are more
scattered around their straight line fit. For the smaller
heliospheres, the BS lies systematically more inward than
predicted from the TS location via the proportionality fit
(\ref{eq-bshp}), and apparently obeys a different linear relation.

For the interstellar ram-pressure dominated models considered
here, the up\-wind-down\-wind asymmetry of the TS is essentially
constant. Excluding the rocket-shape models that have a triple
point in their TS, the relation for the downwind distance $r_{\rm
TSd}$ of the TS is
\begin{equation}\label{eq-tsd}
r_{\rm TSd} = (2.08\pm 0.04) \,\, r_{\rm TS}.
\end{equation}
The appearance of a triple point in the remaining models clearly
yields a different dependence, namely,
\begin{equation}\label{eq-tsdr}
r_{\rm TSd} =
 (1.35\pm 0.12)\,\, r_{\rm TS} + (82\pm 7)\mbox{\rm AU}.
\end{equation}
Again, these fits are shown in Figure \ref{fig-tscorr}, with open
circles representing their corresponding model data. Both subsonic
models \prlocbub\ and \hdsub\ are outliers and were omitted. They
obey a different asymmetry law than the rest of the models,
stemming from the qualitatively different pressure distribution
along the HP in the absence of a bow shock. Model \prone\ is an
outlier with an $r_{\rm TSd}$ smaller by 63 AU than predicted by
the above relation. For small heliospheres, relation
(\ref{eq-tsd}) fits the high density cases \hidensa\ -- \hidensb\
well, but underpredicts the asymmetry of models \hivel\ and
\coldb, which due to their high velocity might be on their way to
the rocket-shaped cases. The mentioned outliers are not included
in the factors of equations (\ref{eq-tsd}) and (\ref{eq-tsdr}).
The pressure distribution in the rocket-shaped cases goes
hand-in-hand with a more inward location of the bow shock; the
models where the BS location is inconsistent with Eq.\
(\ref{eq-bshp}) are the ones where the heliosphere is of that
shape.

In the supersonic solar wind, the total pressure is dominated by
the plasma ram pressure. Since most of the kinetic energy of the
solar wind is converted into heat at the termination shock, it is
effectively the upstream solar wind ram pressure that balances the
total interstellar pressure $P_{\rm tot} = P_{\rm pl} + \rho_{\rm
p} v^2 + P_{\rm H}+ \rho_{\rm H} v^2$, the sum of plasma thermal
and ram pressure and neutral H thermal and ram pressure. Hence a
simple 1-D pressure balance would be achieved, assuming constant
solar wind velocity $v_{SW}$ and an $r^{-2}$ dependence of the
density for the supersonic solar wind region, at the radial
distance of
\begin{equation}\label{eq-pb}
r_{pb}/r_1 = \sqrt{P_1/P_{\rm tot}}
\end{equation}
where $P_{1} = \rho_1 v_{SW}^2$ is the solar wind ram pressure,
and $\rho_1$ the solar wind density, both taken at $r_1$ = 1 AU.
This basic pressure balance distance assumes that all kinetic
energy is converted into heat at the TS. Improving on this by
using the Rankine-Hugoniot relations together with treating the
heliosheath and interstellar flows as incompressible, and assuming
the ISM to be at rest in the heliocentric frame, the TS is
calculated to be at
\begin{equation}\label{eq-TS1}
r_{TS1} = \sqrt{\gamma +3\over 2 (\gamma +1)}\,\, r_{pb} =
\sqrt{7\over 8}\,\, r_{pb}
\end{equation}
\citep{Zank99}, where $\gamma$ is the ratio of specific heats, and
is set to 5/3 in the second equality as well as for all the models
in this paper.

\citet{SuessNerney90} have calculated the case of an ISM that is
moving with respect to the Sun, in which the nose TS position
calculated from equation (\ref{eq-TS1}) is corrected by a weak
dependence on the ratio of interstellar ($v$) and solar wind
velocities,
\begin{eqnarray}\label{eq-TS2}
r_{TS2} &=& \sqrt{2\over \gamma +1}\,\, r_{pb} \,\, \left[
{(\gamma +1)^2 \over 4 \gamma} \left( 1 - {v^2\over v_{SW}^2}
\right) \right] ^{\gamma /2\over\gamma -1} \nonumber \\
        &=& {16\over 5} {1\over\sqrt{3\sqrt{15}}}\,\, r_{pb} \,\,
        \left( 1 - {v^2\over v_{SW}^2} \right) ^{5\over 4} \,\, .
\end{eqnarray}
For $\gamma = 5/3$, the numerical factors of equations
(\ref{eq-TS1}) and (\ref{eq-TS2}) agree with each other to within
0.4\%, and the extra velocity-related factor of (\ref{eq-TS2})
represents a non-negligible reduction in $r_{TS2}$ compared to
$r_{TS1}$ only for high velocity ISM cases, such as model \hivel\
of this paper (a 8\% reduction in this case).

A number of assumptions enter into the derivation of equations
(\ref{eq-TS1}) and (\ref{eq-TS2}). Among them is that these
formulas are only valid for a subsonic ISM, with neutral H absent,
and $P_{\rm tot}$ should be dominated by the ISM thermal plasma
pressure. However, the validity of (\ref{eq-TS1}) and
(\ref{eq-TS2}) can be extended to the supersonic LISM case,
because an interstellar bow shock will convert the flow to
subsonic speeds. It is therefore possible to admit the sum of all
plasma pressure contributions into $P_{\rm tot}$ of equation
(\ref{eq-pb}). If neutral H were tightly coupled to the plasma
(i.e.\ if the mean free paths were very short compared to typical
heliospheric length scales), then the neutral H pressure
contributions can justifiably be included in $P_{\rm tot}$ as
well. However, the neutral-plasma coupling is neither zero nor
very strong, so that the solar wind/LISM pressure balance for the
heliospheres modeled in this paper is more complicated.

The \nummodels\ models show a correlation between $r_{TS2}$ and
the upwind TS location (and consequently, the upwind HP and BS,
equations \ref{eq-hpts} and \ref{eq-bshp}). We present here two of
the $r_{TS2}$ distance estimates. One ($r_{pl}$) is obtained by
using solely the LISM plasma pressure in equations (\ref{eq-pb},
\ref{eq-TS2}) ($P_{\rm tot,pl} = P_{\rm pl} + \rho_{\rm p} v^2$),
effectively neglecting neutral H altogether. The alternative
distance prediction $r_{min}$ uses $P_{\rm tot} = P_{\rm tot,pl} +
P_{\rm H}+ \rho_{\rm H} v^2$, taking the entire pressure
contribution of neutral H into account as well. The correlations
are
\begin{eqnarray}\label{eq-ts2pcorr}
r_{\rm TS} &=& (0.71\pm 0.08) \,\, r_{pl}; \,\,\,
  r_{pl} = r_{TS2}(P_{\rm tot,pl})\\
r_{\rm TS} &=& (1.35\pm 0.02) \,\, r_{min}; \,\,\,
  r_{min} = r_{TS2}(P_{\rm tot})\, .\label{eq-ts2corr}
\end{eqnarray}
The data and the regression fits are shown in Figure
\ref{fig-tscorr}, top, with open diamonds and squares,
respectively. The plasma-only correlation (\ref{eq-ts2pcorr},
diamonds) is poor, and especially pairing models \prtwo\ with
\prone, and \colda\ with \coldb, shows that the same plasma total
pressure does lead to different heliospheres depending on the
neutral contribution. In addition, the high density cases
\hidensa--\hidensb\ are complete outliers (as are \colda\ and
\coldb) where the plasma pressure alone overpredicts the
heliospheric size. For the correlation (\ref{eq-ts2corr})
(squares), it is not necessary to exclude any of the supersonic
heliospheres, in particular not models \hidensa--\hidensb. Only
model \hdsub\ is an outlier, but not model \prlocbub.

The predicted and modeled TS distances in the plasma-only relation
(\ref{eq-ts2pcorr}) scatter around the fit too much, and neutral
dominated models are excluded outright. The deviations from the
predicted locations indicate changes to the pressure balance due
to \ce\ with neutral H. The pressure balance gets shifted inwards
by pickup ion production in the supersonic solar wind that reduces
the supersonic wind speed and hence its ram pressure. On the other
hand, the TS pressure balance can get shifted outwards by other
effects, among them the deceleration of the outer heliosheath
plasma by \ce\ with secondary neutrals. Relation
(\ref{eq-ts2corr}) has a much smaller scatter and therefore is
better suited to predict $r_{\rm TS}$, with the caveat that the
neutral-plasma interaction drives the neutrals out of equilibrium,
and pressure balance can only be described by an empirical factor
of 1.35.

This study is focusing solely on the variation of interstellar
boundary parameters and the heliospheric response. While we do not
wish to study the question of different solar wind conditions on
the heliosphere here, it is interesting to note that all the
results so far only depend on $P_1$, the solar wind ram pressure
at 1 AU, and that the locations of TS, HP, and BS are dictated by
a pressure balance between solar wind and LISM, with empirical
correction factors. In reality, the solar wind is time dependent
on an 11 year cycle and on smaller, episodic timescales. The
variation in solar wind ram pressure leads to small variations in
the heliospheric boundary locations \citep[e.g.,][]{ZankMueller03,
Izmodenov05} that are qualitatively consistent with Eq.\
(\ref{eq-ts2corr}). Similarly, the increased ram pressure in polar
directions from the fast polar wind during solar minimum results
in larger heliospheric distances in these directions as compared
to isotropic slow solar wind \citep{Pauls97,Tanaka99}. An example
of the reaction of the heliosphere to different solar wind ram
pressures while holding the ISM environment constant is the
comparison of models \eind--\eindmodm\ in Table \ref{tbl-res},
where the boundaries move outward for a ram pressure increase
(model \eindmodm), but do not change for a doubling of the solar
wind thermal pressure (model \eindmod).

Recent investigations of astrospheres around other cool
main-sequence stars \citep{Mueller01,Wood02,Wood05b} use a range
of stellar wind ram pressures $P_1$ that are different from the
solar wind, but a modeling strategy identical to the one for the
heliosphere. A cursory analysis of these models (plots not shown
here) shows that equations (1) - (\ref{eq-ts2corr}) hold also for
all these cases within the stated accuracy. This result
underscores the argument of pressure balance, and extends the
results of this section to other ram pressure regimes such as for
astrospheres carved out of the ISM by coronal stellar winds.

\begin{figure}
\epsscale{0.90} \plotone{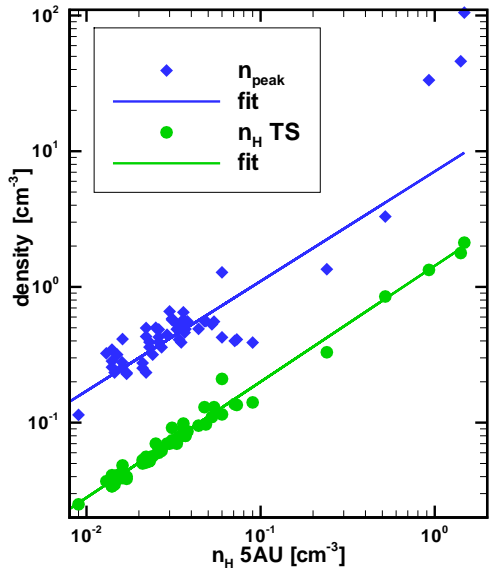} \caption{Correlations between the
neutral H density at 5 AU upwind, and at the TS (circles) and the
peak density inside the hydrogen wall, max($n_H$(wall))
(diamonds).\label{fig-neutcorr}}
\end{figure}

\begin{figure}
\epsscale{0.90} \plotone{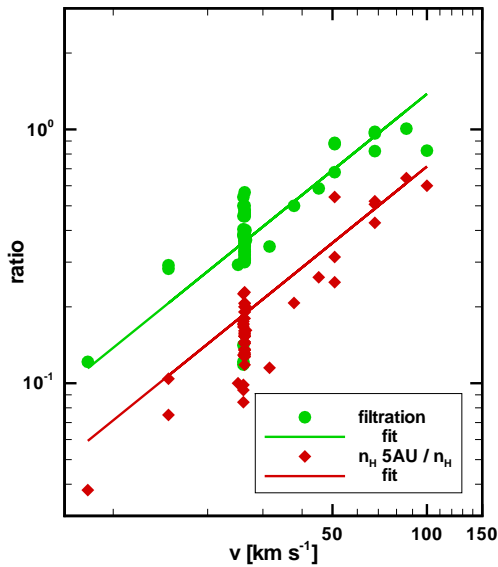} \caption{Correlation between the
filtration ratio $f$=$n_{TS}\mbox{(H)}/n(\mbox{H}^0)$ and the
interstellar velocity $v$ (circles), and correlation of the 5 AU
density ratio $n_{\rm 5AU}\mbox{(H)}/n(\mbox{H}^0)$, with $v$
(diamonds). The solid lines are linear fits while the plot itself
is double-logarithmic.\label{fig-neutvcorr}}
\end{figure}

\subsection{Heliospheric neutral hydrogen\label{ss:neutralcorr}}

The density $n_{TS}\mbox{(H)}$ of neutral hydrogen that crosses
the termination shock at the upwind stagnation axis varies over a
considerable range in the \nummodels\ models, from 0.01 to 0.33
\cc, with four larger values between 0.9 and 2 \cc\ for the high
density models \hidensa--\hidensb\ and \coldb. The filtration
ratio $f$ is the neutral density at the TS divided by the
interstellar neutral density \nH, well upstream of the BS (to
avoid contamination by component 2 and 3 neutrals), and these
relative values vary from 0.1 to 0.6, with a few higher values
0.7--1.0 for higher velocities. The filtration is listed as
$f_{\rm TS}$ in Table \ref{tbl-res} for all \nummodels\ models, as
are similar ratios for the peak hydrogen wall density ($f_{\rm
peak}$) and for the density at 5 AU on the upwind stagnation axis
($f_{\rm 5 AU}$). We choose 5 AU as a fixed reference distance
with the expectation that photoionization is not yet important at
this distance.

The neutral hydrogen at the TS is comprised of original
interstellar material, and of slower secondary neutrals created
upwind of the HP which form the hydrogen wall. After crossing the
HP these neutrals get depleted in the heliosheath by \ce\ which
replaces them with neutrals mostly in outward directions. For high
velocity models, this loss process starts from a higher HP density
so that higher filtration ratios occur (\S \ref{ss:hivel}). Both
the filtration ratios and the absolute TS neutral densities
correlate well with the neutral density $n_{\rm 5AU}\mbox{(H)}$ at
5 AU. The correlations are,
\begin{eqnarray}
n_{TS}\mbox{(H)} & =
  & (1.43\pm 0.02) \,\, n_{\rm 5AU}\mbox{(H)} ^{0.85\pm 0.01}\label{eq-ntsfive} \\
               f & =
  & (1.55\pm 0.13) \,\, \left[ n_{\rm
               5AU}\mbox{(H)}/n(\mbox{H}^0)\right]^{0.72\pm 0.04}\label{eq-ftsfive}
\end{eqnarray}
where all models have been included in (\ref{eq-ntsfive}), and the
high density models \hidensa--\hidensb\ had to be excluded as
outliers in (\ref{eq-ftsfive}) (plot not shown). The data and the
fit (\ref{eq-ntsfive}) are displayed in Figure \ref{fig-neutcorr}
by circle symbols. This correlation is interesting in that it
relates a density at a fixed distance to a density at the TS
regardless of the actual distance of the TS. The net loss in the
region of the supersonic solar wind is not sensitive to the length
of the neutral particle trajectories between the TS and 5 AU
because most of the \ce\ relevant to the 5 AU density takes place
immediately upwind of that distance, given that the local plasma
density is larger with smaller distance, resulting in smaller \ce\
mean free path lengths.

The peak densities in the hydrogen wall range from 1.1 \nH\ to
massive hydrogen walls of 7 \nH, corresponding 0.09--105 \cc\ in
absolute units. The absolute peak density is weakly correlated to
the 5 AU value, as demonstrated by the diamond symbols in Figure
\ref{fig-neutcorr}. Even on a log-log plot, however, a large
scatter of the results around possible power laws is evident.
According to a regression analysis that omits the high density
models \hidensa--\hidensb,
\begin{equation}\label{eq-fivepeakcorr}
n_{peak}\mbox{(H)} =
(7.1\pm 1.5) \,\, n_{\rm5AU}\mbox{(H)}^{0.81\pm 0.06} \, .
\end{equation}
Nonetheless, the fit is poor. The same problem is encountered when
relating the neutral results to the interstellar velocity. There
is a general trend of higher neutral densities with higher
velocity. Figure \ref{fig-neutvcorr} shows the filtration results
(circles) and the relative densities at 5 AU (diamonds) plotted
against the interstellar velocity. It is evident that any
correlation derived from these model results is not unique, as
there are many data points around $v \approx 26$ \kms\ whose
normalized neutral results vary by a factor of 2.7. The fit lines
shown in Figure \ref{fig-neutvcorr} are $f =$ $v$/($73\pm 7$ \kms)
and $n_{\rm 5AU}\mbox{(H)} / n(\mbox{H}^0) = v$/($140\pm 8$ \kms).

By way of a cautionary note, it should be noted that the above
relations are likely model dependent, in that using another
self-consistent modeling strategy for the heliosphere, such as a
particle kinetic model for the neutral H
\citep{Baranov1993,Mueller00,Heerikhuisen06}, might result in
different coefficients for the relations
(\ref{eq-hpts})--(\ref{eq-tsdr}), and
(\ref{eq-ts2pcorr})--(\ref{eq-fivepeakcorr}). Short of carrying
out a parameter study with these alternate models and comparing
the results, it is impossible to incorporate this systematic error
into the error estimates of the coefficients given in the above
relations.

\section{DISCUSSION \label{ss:discuss}}

\subsection{Heliospheric Morphology and Neutrals}

We are now in a position to discuss some of the consequences that
different interstellar environments have for the solar system and
for Earth. There are several models listed in Table \ref{tbl-res}
representing heliospheres that are so small that the outer planets
find themselves beyond the supersonic region of the solar wind, at
least for parts of their orbits. As examples, Figures
\ref{fig-2d20} and \ref{fig-2dcyg} have the orbits for the gas
giants Saturn, Uranus, and Neptune marked in them. An assumption
in these plots is that the ISM flow vector is close to the plane
of the ecliptic, as is the case in the current environment. For
larger angles, the ecliptic TS distances are farther due to the
upwind/downwind asymmetry of the heliosphere.

In the high density case \hidensb\ (Figure \ref{fig-2d20}), Uranus
periodically crosses the TS, traverses the hot inner heliosheath,
crosses the HP, and spends part of its orbit in the shocked LISM
(the outer heliosheath), before crossing the HP again in a reverse
course of events. Neptune is always surrounded by hot, shocked
plasma (either the solar wind heliosheath and heliotail or the
interstellar plasma of the outer heliosheath), and is never
upstream of the TS. Similarly, in the high velocity case of model
\cyg\ (Figure \ref{fig-2dcyg}), part of Neptune's orbit is in the
subsonic solar wind of the heliosheath, and Uranus periodically
comes close to the TS.

For planets in the hot inner heliosheath it can be expected that
solar wind injection into planetary magnetospheres is more
efficient due to increased thermal plasma velocity and solar wind
turbulence. Also the solar wind magnetic field strength is
enhanced, and an increased flux in both galactic cosmic rays and
heliospheric energetic particles is to be expected. Planets that
cross into the outer heliosheath will be exposed to not quite as
hot a plasma, carrying a basically undetermined magnetic field.
However, the planet will then be exposed to the increased neutral
density of the hydrogen wall. This neutral flow will strike the
planetary atmosphere unimpeded and lead to atmospheric drag and
other effects \citep{Yeghikyan04terr}.

None of the \nummodels\ models considered here leads to such dire
predictions for Earth's orbit, and extrapolating from equations
(\ref{eq-TS2}) and (\ref{eq-ts2corr}) beyond their validity,
interstellar densities of 1500 \cc\ with moderate interstellar
velocities, or a velocity of 345 \kms\ with moderate interstellar
densities, would be needed to place the TS at 1 AU. It should be
expected that both these numbers are only a qualitative estimate,
and that detailed modeling of such heliospheres would have to take
into account additional physical processes
\citep{Yeghikyan03,Yeghikyan04}.


As many models in this paper have an inner boundary at or beyond 1
AU, we want to focus on the reference distance of 5 AU instead to
assess the change in particle environment that occurs with
different Galactic environments under the assumption of an
unchanging solar wind. At this distance, the solar wind is
supersonic for all \nummodels\ models. The ratio of interstellar
(slow) neutral hydrogen to solar wind protons at 5 AU upwind is
between 7\% and 25\% for the contemporary heliosphere (models
\preight--\azz). As already mentioned in \S\ref{ss:neutralcorr},
this density ratio scales with interstellar velocity, ranging from
from 2\% to 37\% (models \hdsub\ and \xiboo-\eindmodm,
respectively; model \hivel: 120\%, model \cyg: 45\%), and is
highest for the high density models \hivel--\hidensb\ and \coldb.
In particular, there is more than seven times more neutral H than
solar wind protons at 5 AU in models \hidensc\ and \hidensb.

Naturally, the increased presence of interstellar \HI\ increases
the rate of \ce\ and hence the production rate of fast component 3
neutrals, the so-called neutral solar wind (NSW). The NSW density
at 5 AU for the contemporary heliospheres is $\approx $4 $\times
10^{-4}$ \cc, but increases to 0.014, 0.018, and 0.023 \cc\ for
models \hidensa, \hidensc, and \hidensb, respectively. A similar
drastic relative NSW increase should occur at Earth orbit as well
for these high density cases. In contrast, the high velocity case
\cyg\ only yields a five-fold increase of NSW to 0.002 \cc\ at 5
AU.


\subsection{Cosmic ray transport model}


As an application of the results discussed above we make use of a
cosmic ray transport model for three demonstrative solutions of
GCR phase space density. They are calculated for the interstellar
environments corresponding to the Local Bubble, the LIC, and a
dense cloud of mostly neutral hydrogen, based approximately on
models \prlocbub, \acen, and \hidensa, respectively. The
cosmic-ray transport model we use is discussed in detail in
\citet{FlorinskiZank05} and the reader is referred to that paper
for a complete description. Briefly, the plasma flow background
obtained from the multi-fluid code is used to calculate all 3
components of the heliospheric magnetic field in the azimuthal
plane from Faraday's law combined with the zero divergence
condition with a Parker spiral field specified at the inner
boundary \citep{Florinskietal03b}. The modified field component of
\citet{JokipiiKota89} that alters the transport parameters at high
heliographic latitude is also included in the model. The
interstellar magnetic field plays no role in this model because it
fluctuates on scales that are typically much larger than diameter
of the cosmic ray gyroorbit and hence has little effect on the
particle's trajectories. Next, magnetic turbulent energy
$\langle\delta B^2\rangle$ and the associated turbulence
correlation length in the solar wind region are computed from the
hydrodynamic model of incompressible turbulence transport
\citep{Zanketal96,Matthaeusetal99}. The model assumes that the
number of waves propagating parallel and antiparallel to the mean
magnetic field are equal and ignores certain wave propagation
effects by neglecting the Alfv\'en speed compared with the mean
plasma velocity in the solar wind.

Little is known observationally about the turbulent content of the
inner heliosheath, and turbulence transport in that region is
poorly understood at present. Here we use a simple assumption that
the turbulent ratio $\langle\delta B^2\rangle/B^2$ and the
correlation length are both constant across the termination shock
and in the heliosheath. This assumption is based on the physics of
Alfv\'en wave transmission through a quasi-perpendicular shock
that yields the expression relating the turbulent ratio on the two
sides of the shock as \citep{McKenzieWestphal69}
\begin{equation}
\label{alfwtrans} \frac{\langle\delta B_2^2\rangle}{B_2^2}
=\frac{(s+1)}{2s}\frac{\langle\delta B_1^2\rangle}{B_1^2},
\end{equation}
which is not too different from 1 for shocks of moderate strength
(compression ratios $s =$ 2.5--3.0).

The turbulent content of the solar wind is strongly influenced by
the process of Alfv\'en wave generation by pickup ions as they
scatter from the initial ring-beam distribution onto a bisphere in
velocity space \citep{WilliamsZank94, Isenbergetal03}. Because
pickup ions are produced in charge transfer collisions between
solar wind protons and interstellar hydrogen atoms, changes in
neutral density are the principal source of turbulence
variability. The latter translates into variations in the amount
of GCR modulation through an appropriate diffusion model. We use
the Quasi-Linear theory (QLT) \citep[e.g.,][]{Jokipii66} for the
parallel component of the diffusion tensor and the Nonlinear
Guiding Center (NLGC) theory \citep{Matthaeusetal03} for the
perpendicular component. The former is governed by the
fluctuations with wavevectors oriented parallel to the mean
magnetic field (the slab component), while the latter is
determined by the fluctuations orthogonal to the field (the 2D
component). We assume that slab fluctuations comprise 10\% of the
total energy with 2D making up the rest, which is in agreement
with the observed ratio \citep{Bieberetal96}.

Both QLT and NLGC require a specification of the reduced
(one-dimensional) turbulent spectrum. The power spectrum measured
in the solar wind at low latitudes consists of a flat energy range
followed by a Kolmogorov inertial range with $k^{-5/3}$
\citep[e.g.,][]{Bieberetal94}, while some observational evidence
points to a $k^{-1}$ dependence at high latitudes
\citep{HorburyBalogh01}. The low end of the turbulence spectrum is
likely to be populated by structures (shocks and discontinuities)
that may be responsible for the modulation of very high energy
(above 1 GeV) particles. Accordingly, we use two forms of the
power spectrum, identical in the inertial range, but having a
different spectral index in the energy range: 0 (diffusion Model
I) and $-1$ (diffusion model II). As discussed in
\citet{FlorinskiZank05}, the first model emphasizes modulation in
the heliosheath region, while model II is dominated by solar wind
modulation. Because the amount of modulation in the heliosheath is
not known at present we use two plausible models to cover the full
range of possibilities.

\subsection{Cosmic ray modulation results}


\begin{figure}
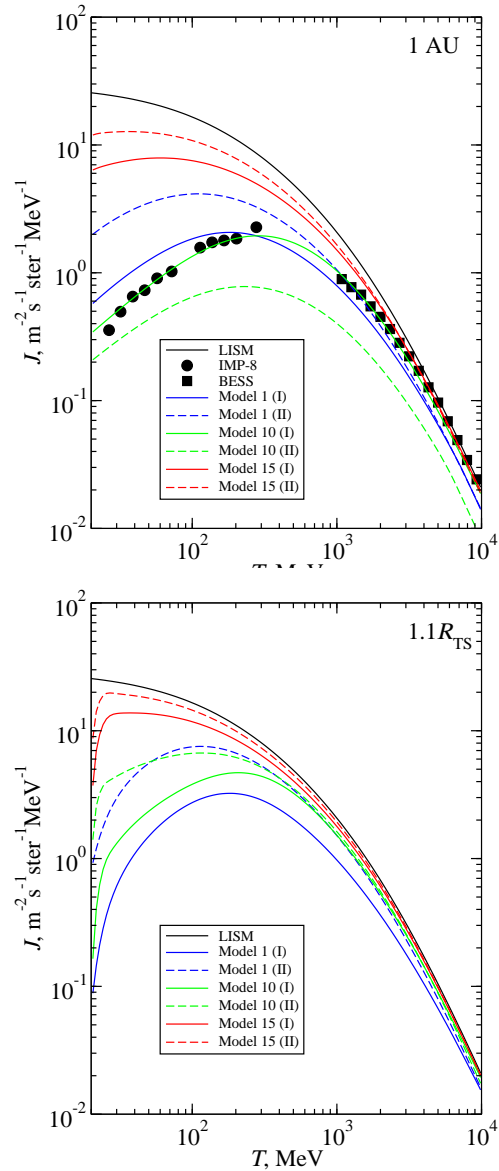

\epsscale{0.85} \plotone{c9a.eps} \plotone{c9b.eps} \caption{GCR
proton differential spectra at 1 AU (top) and just beyond the
termination shock (bottom) in the apex direction for the three
interstellar environments represented by models \prlocbub, \acen,
and \hidensa. The ISM GCR spectrum assumed at the external
boundary \citep{IpAxford85} is shown with a solid black line. The
roman numerals following the model number refer to turbulence
evolution models I and II. 1 AU spectra observed during the time
around the 1995 solar minimum are shown for comparison. The low
energy data are from IMP-8 satellite observations
\citep{McDonald98} while high energy data are from BESS balloon
measurements \citep{Sanukietal00}. \label{fig-gcr}}
\end{figure}


The different sizes of the heliosphere as well as the different
particle distributions will affect the modulation of GCRs as they
pass from the ISM through the heliosphere into interplanetary
space and to Earth. The GCR transport coefficients depend on the
heliospheric magnetic field as well as on the level of plasma
turbulence. While the magnetic field geometry, its strength, and
the properties of solar wind turbulence (especially in the
heliosheath) are far from having been studied conclusively,
theoretical models of cosmic-ray transport can be constructed on
the basis of known physics
\citep{Florinskietal03b,FlorinskiZank05}. Here we perform a series
of computer simulations of GCR modulation in the global
heliosphere for models \prlocbub\ (heliosphere in the Local
Bubble), \acen\ (contemporary heliosphere), and \hidensa\ (a high
density cloud encounter).

A thicker heliosheath such as in model \prlocbub\ could, in
principle, be expected to yield stronger modulation of GCRs (i.e.\
a lower particle flux arriving at Earth). However, the level of
magnetic field turbulence production, which in turn depends partly
on PUI production, can be less efficient in low density
environments. Figure \ref{fig-gcr} shows the high energy proton
GCR spectra for the three cases (models \prlocbub, \acen, and
\hidensa) calculated with diffusion models I and II. The right
panel of this figure shows particle intensity at the termination
shock thus demonstrating the effect of heliosheath modulation.
Because low energy GCR protons typically do not reach 1 AU, we
focus on the high energy end of the spectra.

It follows from our model calculations that the GCR environment at
Earth for the Local Bubble scenario (model \prlocbub) would have
been less intense than at present if diffusion model I was
correct, or, surprisingly, more intense with model II diffusion
coefficients. The latter is a consequence of the large GCR mean
free path predicted by the second diffusion model for the solar
wind region in model \prlocbub\ in the absence of PUI turbulence
driving. In both cases the cumulative heliosheath GCR modulation
is more important for the Local Bubble environment than for the
contemporary heliosphere. This is a combined effect of a stronger
TS, a larger decrease in the radial mean free path across the
shock, and a thicker heliosheath.

For an encounter with a high density cloud (model \hidensa), the
heliospheric GCR shielding is much less effective than in the
contemporary heliosphere, such that the high energy part of the
GCR spectrum approaches the one assumed for the pristine LISM.
Here, the opposing effects of a relatively large heliosheath
diffusion coefficient combined with a much smaller extent of the
modulation cavity, and enhanced PUI turbulence driving in the
solar wind result in a significant reduction in the heliospheric
shielding of GCRs. The predicted cosmic-ray intensity increase at
Earth is between 1.4--2.4 (model I) and 4.1--7.6 (model II) in the
energy interval between 300 MeV and 1 GeV.

In addition to GCRs, the distribution of anomalous cosmic rays
which are accelerated at the TS, will depend on the background
neutral density, and the strength and distance of the TS
\citep{Florinskietal03a}. Together, the cosmic ray environment at
Earth influences the terrestrial magnetosphere as well as climate,
atmosphere, and biology
\citep[e.g.][]{Scherer00,Yeghikyan04terr,Frischetal:2002}.

\section{CONCLUSIONS}

On its path through the galaxy, the Sun has encountered (and will
encounter) different interstellar environments. This motivates a
parameter study to investigate the response of the heliosphere to
these changing conditions under the assumption of a constant solar
wind. For conditions that are not too far from the contemporary
LISM environment, the following findings emerge from analyzing
\nummodels\ self-consistent multi-fluid models.

\begin{enumerate}

\item  Allowing generous assumptions about the LIC morphology, the
LIC column density towards nearby stars indicates the Sun first
encountered the LIC gas within the past 40,000 yr, and the CLIC
within the past $\sim$60,000/$\tilde{f}$ yr (where $\tilde{f}$
represents the fraction of space filled by the CLIC).  The Sun is
expected to exit the LISM gas cloudlet, which is characterized by
the common LIC velocity, sometime within the next $\sim$0--4000
yr. In general, passage through interstellar clouds will lead to
variations in the heliosphere boundary conditions over timescales
possibly as short as 10$^3$ yr. Nearby ISM generally resembles low
column density ISM observed elsewhere.

\item The size of the heliosphere is determined by the balance of
solar wind and interstellar pressure. For the investigated
parameter range, in which the LISM is mostly ram-pressure
dominated, the upwind termination shock distance can be estimated
by equation (\ref{eq-ts2corr}), using equations (\ref{eq-pb}) and
(\ref{eq-TS2}). This relation is derived from a pressure balance
argument modified by an empirical factor expressing the efficiency
of the neutral pressure contribution to the overall interstellar
pressure.

\item Heliocentric distances of interest such as the heliopause,
the bow shock, or the upwind and downwind termination shock scale
linearly with each other (e.g.\ $ r_{\rm HP} = 1.39 \,\, r_{\rm
TS}$, $r_{\rm BS} = 1.95 \,\, r_{\rm HP}$). Therefore, when the
upwind termination shock distance is predicted in an absolute way
as above, the other distances can be predicted as well. However,
the scalability and predictability of the heliosphere size with
these relations are only applicable to parameter sets in which the
LISM flow is supersonic. The subsonic cases, when the Sun is
surrounded by hot plasma or alternately when the Sun and the
surrounding interstellar cloud are comoving in space, obey a
different set of correlations, and are generally more difficult to
model numerically.

\item For low interstellar velocities, the heliosheath and
heliotail plasma are subsonic throughout, and the ratio of
downwind to upwind termination shock distance (TS asymmetry) is
2.1. For higher velocities, the heliosphere assumes a rocket
shape, with a modified pressure balance in the downwind
directions.

\item Neutral hydrogen results such as the filtration ratio, the
peak hydrogen wall density, or the density at 5 AU upwind of the
Sun, correlate with each other. Their absolute value is weakly
correlated to the interstellar velocity with which the neutrals
arrive at the respective heliosphere, as the charge exchange mean
free path depends on this velocity, and higher velocities shorten
the heliocentric distances to the heliospheric boundaries.

\item For encounters with a high density interstellar cloud
($\sim$15 \cc, about 50 times the contemporary value), the
particle fluxes arriving at Earth orbit, including interstellar
neutrals, neutral solar wind, and cosmic rays will increase
markedly. These changes potentially affect Earth's atmosphere and
its climate. The changes in particle fluxes just due to a higher
interstellar velocity are less pronounced.

\item For the period when the Sun was embedded in the Local
Bubble, particle fluxes were reduced substantially. Secondary
particles like anomalous cosmic rays and neutral solar wind were
entirely absent, and the galactic cosmic ray flux arriving at
Earth was comparable to the contemporary flux, or even reduced,
depending on the modulation model.

\end{enumerate}

\acknowledgments

This material is based upon work supported by the National
Aeronautics and Space Administration (NASA) under grants issued
through the Science Mission Directorate, namely through the
Astrophysics Theory Program under grant NAG5-13611, and through
the Solar and Heliospheric Physics grant NAG5-12879. PCF would
like to thank NASA for support through grants NAG5-13107 and
NAG5-11005 to the University of Chicago. GPZ, VF and HRM
acknowledge partial support from an NSF-DOE grant ATM-0296114 and
NASA grants NAG5-12903 and NAG5-11621.


\bibliographystyle{apj}
\hyphenation{Post-Script Sprin-ger}

\end{document}